\documentclass[12pt]{article}
\usepackage[english]{babel}
\usepackage[utf8x]{inputenc}
\usepackage{amsmath}
\usepackage{amssymb}
\usepackage{mathtools}
\usepackage{graphicx}
\usepackage{multicol}
\usepackage[margin=1in]{geometry}
\usepackage{times}
\usepackage{setspace}
\usepackage{amsthm}
\usepackage{mathtools}
\usepackage{natbib} % APA style

\theoremstyle{definition}% upright text, extra space above and below

\edef\hc{\string:} % \hc prints a normal colon

\title{Curricular Analytics: A Framework for Quantifying the Impact of Curricular Reforms and Pedagogical Innovations\thanks{This paper is a prerelease created for the purpose of obtaining feedback. To properly cite this work, please contact the corresponding author in order to obtain the most up-to-date information.}} 
\author{ Gregory L. Heileman$^{\dagger\natural}$ \and Chaouki T. Abdallah$^{\ddagger}$ \and Ahmad Slim$^{\ddagger}$ \and Michael Hickman$^{\ddagger}$ \\~\\
  greg.heileman@uky.edu, \{chaouki, ahslim, mshickman91\}@unm.edu \\~\\
  $^\dagger$Department of Electrical \& Computer Engineering \\
  University of Kentucky \\ \\ 
  $^\ddagger$Department of Electrical \& Computer Engineering \\
  University of New Mexico \\ \\
  $^\natural$corresponding author }

\date{}

\begin{document}

\maketitle
%\newpage

%----------------------------------------------------------------------------------------
%\doublespacing

\begin{abstract}
In this paper we articulate a framework for quantifying the complexity of curricula based on their fundamental structural and instructional properties. We then introduce the notion of \emph{curricular analytics} as a means of relating curricular complexity to student success outcomes, and we demonstrate the usefulness of curricular analytics by applying them to a number of important problems.
\end{abstract}

{\bf keywords:} curriculum, curricular analytics, student success, curricular complexity

%------------------------------------------------
\section{Introduction}
In this paper we investigate analytics related to the most fundamental element of student success, namely the curriculum pathways that students attempt to follow on the way to attaining learning outcomes, and eventually earning their degrees. In general, the promise of analytics in higher education is that they can be used to inform decision-making in ways that improve student success outcomes. For instance, there are numerous efforts where student demographics and prior student performance have been used to direct interventions~(e.g., counseling, mentoring, tutoring, etc.) aimed at improving retention and graduation rates~(see \cite{DiHu:13,MaDiMa:16,PiAr:10}). Our contention, however, is that with regards to student success outcomes, progression within a curriculum is the foundation of student academic success. If students are hindered in any way and for any reason while trying to follow their intended curricular pathways, they may be delayed in graduating, and their risk of stopping out of school increases. Indeed, we can~(in fact, should) view most student success interventions in terms of their impact on progression towards the degree. Thus, it makes sense to study these interventions at the most basic level by considering their direct impact on student progression.  

Philosophically, our work can be thought of as a reductionist approach to the study of student success, akin to how those in the natural sciences often explain biological phenomena in terms of the underlying chemistry, which in term might be explained more fundamentally using the laws of physics. The problems with this approach are myriad. First, there is the inherent difficulty associated with quantifying the impact a particular intervention or reform has on a given student's progress within a given degree program. Consider the difficulty of quantifying the impact that an internship program might have on student progression in all of the academic programs associated with those students participating in internships. This is often exacerbated by the fact that in many cases those implementing interventions do not have the opportunity to coordinate with those responsible for providing the curriculum. For instance, it is not uncommon for student affairs staff to provide services to students aimed at increasing student success, without being provided the opportunity to discuss these services with faculty in all of the programs they impact. In this case, the implementors of the interventions may not have a solid grasp of curriculum confronting the students they hope to impact, and faculty may not fully appreciate how these services could further the attainment of the learning outcomes associated with the curriculum. This is just one of the consequences of operating in a manner that some characterize as ``silos.'' 

Second, the shared governance model practiced on most campuses can lead to a further hardening of these silos. The mantra often stated when the subject of curriculum is broached is, ``faculty own curriculum,'' taken by many to mean ``stay out of the faculty's business.'' In some cases, faculty are fully justified in this aggressive defense of their territory. Well-meaning administrators, perhaps spurred on by trustees or state legislators, often ask programs to change their curricula~(e.g., reduce total credit hours) as an end in and of itself. That is, the goal appears to be the curriculum change, rather than its ability to improve program quality, further the attainment of student learning outcomes, or enhance some other measure of student success. 

Curricular changes driven by metrics that demonstrate student benefits are far more likely to be endorsed by faculty than those emanating as top-down dictates from administration. Indeed, curriculum committees commonly approve curricular reforms, and faculty continuously experiment with pedagogical innovations, in efforts to improve student progress. These curricular or pedagogical experiments are often ones that have been reported to work at other institutions, and the hope is that they will also prove effective when implemented in a new environment. In essence, without a reliable means for predicting expected efficacy, these student success experiments amount to ``shots in the dark.'' In our experiences, a typical outcome is anecdotal evidence of how students and/or faculty felt about the intervention, along with how much or little they believe it helped. 

Finally, there is the problem of determining the best locations within a curriculum where interventions should be focused. For many curricula there is a belief---often informed by experience, intuition or folklore---that some courses are more important than others. This notion of ``importance'' is bestowed upon courses for numerous reasons. For instance, earning a high grade in a particular course may be a strong predictor of future success in the curriculum as a whole, there may be a low-pass-rate course that serves as a gateway to many other courses in the curriculum~(a course that acts as a filter), or there may be a course that is foundational to a discipline which many other courses in the curriculum build upon during the junior and senior years. We contend that the complexity of a curriculum is directly related to the number of such important courses it contains. In this paper we describe methodologies for more formally quantifying the notion of important courses, and we use them to characterize the complexity of a curriculum as a whole.

A more formal analytical framework built around the notion of measuring the extent to which curricular or pedagogical interventions impact student progression is needed. In this paper we propose such a framework, and we describe some of the benefits that entail from its use. In particular, we have used this framework to foster analytics-based discussions with faculty and curriculum committees---initiating these conversations with data and a systematic framework, rather than with opinion and ill-defined success criteria. We have found that this approach often reduces the emotion that accompanies opinion-based reform discussions, thereby promoting the opportunities for consensus formation around proposed reforms. It is also worth noting that by creating meaningful metrics related to the structure of curricula, we provide a new and effective means for comparing and contrasting curricula.  We have seen faculty make good use of curricular metrics in order to compare their programs to those offered by their aspirational peers, leading to data-informed decision-making around curriculum reform.

In addition to enhancing opportunities for meaningful engagement with faculty, a curricular analytics framework provides a means for modeling a given educational environment, and for predicting the likely impact of particular curricular and pedagogical reforms within that environment. That is, it supports a more formal approach for planning and evaluating student success reforms, providing a cornerstone for predictive analytics related to curriculum. Thus, the promise of curricular analytics is that by measuring the direct impact of interventions on curricular progression, we can more directly tie particular interventions to student success outcomes. 

% anon edit: In our work, we have been proponents of the importance
We are proponents of the importance of recognizing that reforms occur within a larger educational context that must be properly understood and characterized if one hopes to optimize the impact of particular interventions. This approach treats the university as a complex system, comprised of a set of sub-components that interact in order to create the system as a whole, with each component contributing in some manner, either directly or indirectly, to the success of improvement efforts~\citep{AbBaHe:16,AbHeBa:17,Ro:16}. An important point to note is that from one university to the next, the properties of the university system will differ. Ideally, before launching improvement initiatives at great effort and expense, one would determine the most important factors that contribute to attrition and persistence, and would use these to construct a model that can be used to accurately predict the expected improvements that can be obtained by implementing specific reforms at particular universities. This requires a formalization of the university system model we have just described.

A difficult aspect of the systems formulation of a university involves determining the proper metrics and measurements that can be used to quantify the various sub-components of the educational system. This is necessary if we hope to analyze the system in order make predictions, quantify the impact of interventions, and focus efforts where they are most likely to succeed. Curricular analytics contributes to this systems-theoretic way of thinking about educational reform. Specifically, we consider formal ways of characterizing the inherent complexity of university curricula, and through simulation we then demonstrate a direct relationships between the complexity of curricula, and the success of students attempting to navigate through curricula to graduation.

% anon edit: In this paper we summarize our work in the area curricular analytics over the past few years~\cite{HeHiSlAb:17,Hi:17,Sl:16,Wi:14,WiHeSlAb:14},
In this paper we summarize recent work related to curricular analytics~\citep{HeHiSlAb:17,Hi:17,Sl:16,Wi:14,WiHeSlAb:14}, and we articulate a new framework that organizes this work in a manner that facilitates practical applications and supports further theoretical development of curricular analytics. An overview of the paper is as follows. In Section~\ref{theory} we develop a background for the study of curriculum by decomposing it along lines related to student success outcomes. In Section~\ref{practice} we show how to use this decomposition as a part of a framework that allows us to quantify the impact of various curriculum-related factors on student success. Next, in Section~\ref{application} we provide a few examples of how to apply this framework, and the analytics it generates, to support curriculum-based improvement efforts. Finally, in Section~\ref{conclusion} we provide directions for the further development of curricular analytics in higher education.

%------------------------------------------------
\section{Curricular Analytics---Theory}\label{theory}
\begin{figure}
 \centerline{\includegraphics[width=6.5in]{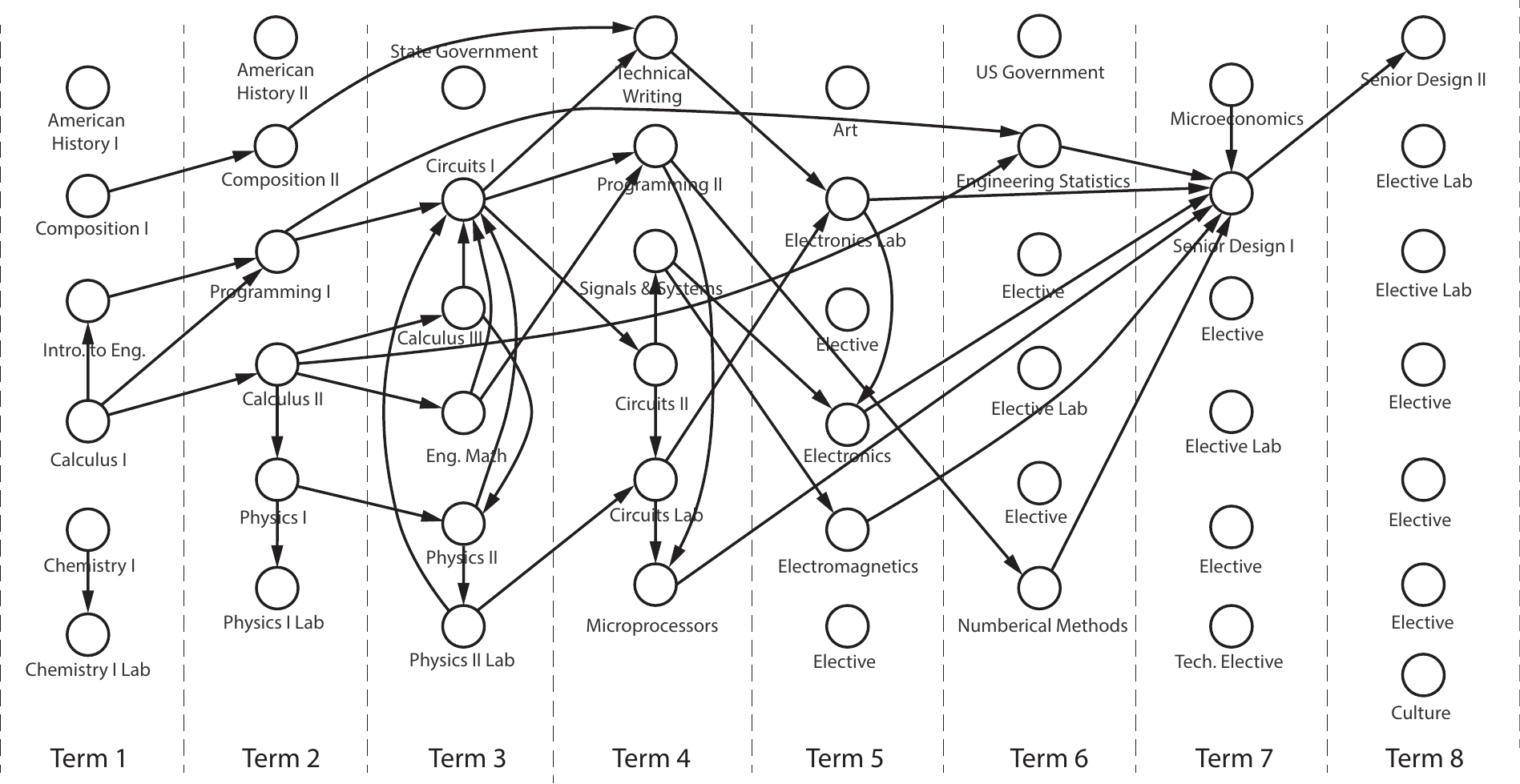}}
 \begin{center}{\bf (a)}\end{center}~\\~\\
 \centerline{\includegraphics[width=6.5in]{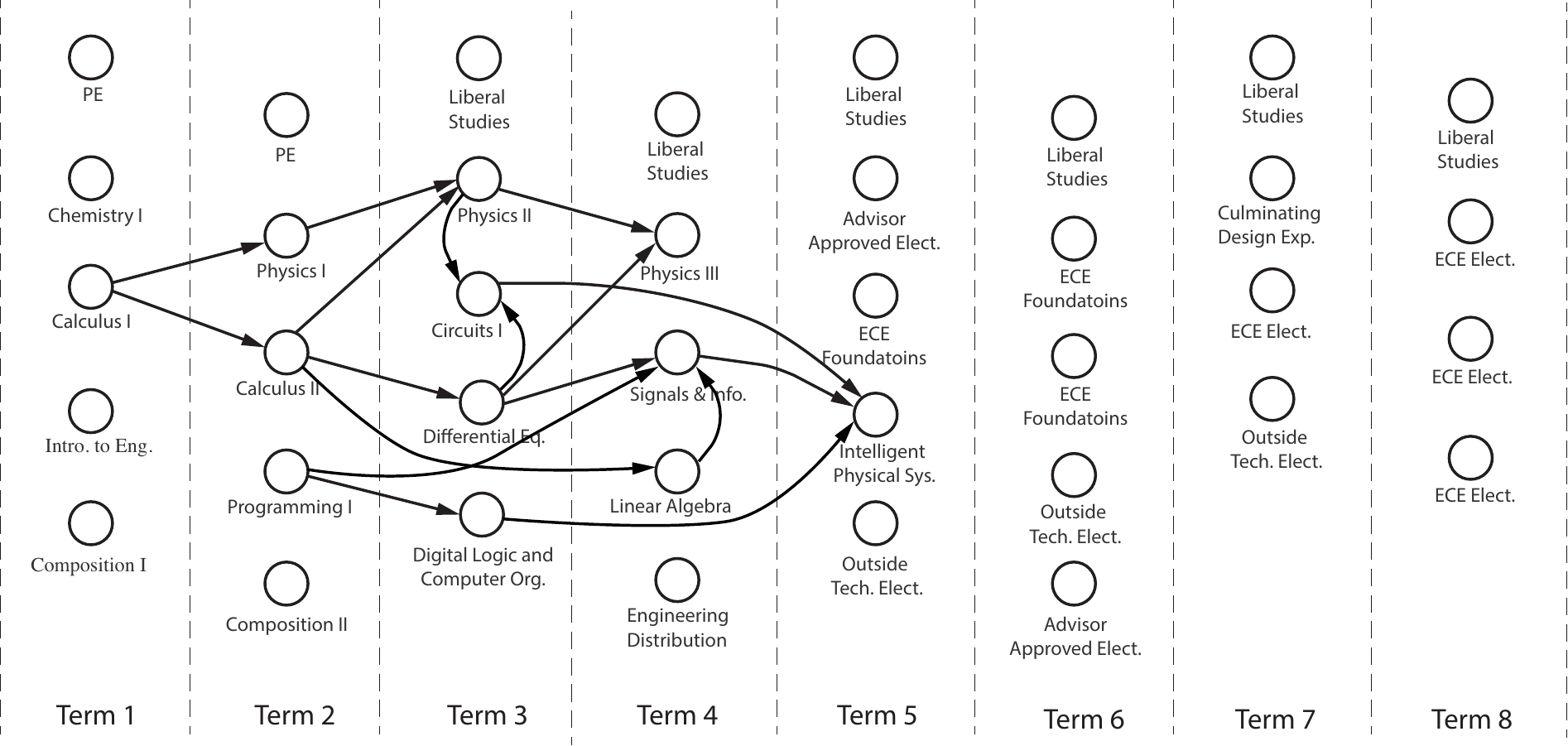}}
 \begin{center}{\bf (b)}\end{center}
 \caption{The four-year degree plans for electrical engineering programs at two large public institutions holding identical ABET accreditation.}\label{univ-curric}
\end{figure}

The analysis of curricula is greatly facilitated by the fact that most academic programs publish their curricula on public websites. This allows those in one academic program to compare and contrast their curriculum to the curricula offered by other similar programs.  For instance, in Figure~\ref{univ-curric} we show the undergraduate electrical engineering curricula provided by two large public institutions in the United States that are similarly accredited by ABET~(ABET, 2017).\nocite{ABET:2017-18}
The term-by-term organization of these curricula constitute the four-year (eight-term) \emph{degree plans} that students are expected to follow. We have drawn these curricula as graphs, where the vertices represent courses, and the directed edges represent \emph{prerequisite} arrangements between courses. That is, for each directed edge, the source vertex is a prerequisite course that must be passed prior to taking the course associated with the destination vertex. If there is a directed edge between two courses in the same term, the source vertex is a \emph{co-requisite} course that may be taken either prior to or at the same time as the course associated with the destination vertex. If two courses must be taken together in the same term, we refer to them as \emph{strict co-requisites}. The direction of the edge between strict co-requisite courses is irrelevant. 

It is interesting to note that the two programs shown in Figure~\ref{univ-curric} have identical ABET accreditation. This means that each program satisfies the same eleven ABET program learning outcomes. Thus, from the perspective of ABET, each program is of sufficient quality that the engineers they produce should be prepared to have successful careers. Even though these programs are identically accredited with identical program learning outcomes, it is readily apparent that their structures are quite dissimilar. In particular, the curriculum in Figure~\ref{univ-curric} (a) appears far more complex than the curriculum in Figure~\ref{univ-curric}~(b). Students attempting the former must satisfy a much larger number of pre- and co-requisite constraints than those attempting the latter. A logical question that arises is, can we quantify these differences in a manner that leads to useful analytical results? For instance, what is the expected graduation rate for similarly prepared students in each curriculum? A host of additional questions come to mind when comparing these two curricula. For example, what is the most important course in each curriculum, and by how much would the success rates of students attempting to complete these curricula improve with small improvements in these courses? Perhaps the most important question is, does one of these programs better prepare a student for success in their chosen field than the other? In the next two sections we provide a framework and toolset that allows curriculum designers to investigate these questions in more detail, and answer them under reasonable assumptions. 

%------------------------------------------------
\subsection{A Framework for Analyzing Curriculum}\label{framework}
The challenge associated with studying the impact of curricula on student progression involves finding a useful decomposition of the various curriculum-related components that influence this progress. We demonstrate the usefulness of modeling the overall complexity of a curriculum as a function of two main components, (1) the manner in which courses in the curriculum are taught and supported, and (2) the manner in which the curriculum is structured. We refer to the former as the \emph{instructional complexity} of the curriculum, and to the latter as the \emph{structural complexity} of the curriculum. Each of these main components are functions of numerous other curriculum-related factors. 

More formally, let $\bar{x}_c$ denote a vector of measurements consisting of all factors that characterize the instruction associated with a given curriculum $c$. These factors quantify items such as the quality of the instructors who teach the courses in the curriculum, the availability of course offerings, the support services~(e.g., tutoring, supplemental instruction, etc.) provided, the inherent difficulty of course topics, etc. The instructional complexity component of curriculum $c$, denoted $\gamma_c$, is defined as a function, $g$, of these factors:
\begin{equation}
 \gamma_c = g(\bar{x}_c).
 \label{instruct_complex}
\end{equation}

Next, let $\bar{y}_c$ denote a vector of measurements consisting of all factors that characterize the structural properties of curriculum $c$. These factors include the total effort~(i.e., credit hours) associated with the curriculum, the manner in which the term-by-term courses within the curriculum are organized, the pre- and co-requisite relationships within the curriculum, etc. The structural complexity component of curriculum $c$, denoted $\alpha_c$, is defined as a function, $h$, of these factors:
\begin{equation}
 \alpha_c = h(\bar{y}_c).
 \label{struct_complex}
\end{equation}

Finally, the complexity of curriculum $c$, referred to as \emph{curricular complexity} and denoted $\Psi_c$, is defined as a function, $f$, of the structural and instructional complexity components:
\begin{equation}
 \Psi_c = f(\alpha_c, \gamma_c). 
 \label{curric_complex}
\end{equation}

The remainder of this paper is concerned with characterizing $\Psi_c$, $\alpha_c$ and $\gamma_c$, and using these characterizations to create useful analytics that can be used to guide curricular and instructional reform discussions. This work is complicated by the fact that it is difficult to quantify and measure all of the factors in $\bar{x}_c$ and $\bar{y}_c$ that influence structural and instructional complexity, making it difficult, perhaps impossible, to fully characterize the exact functional forms of $f, g$ and $h$. However, by making a few reasonable assumptions about the relationships between $\Psi_c$, $\alpha_c$ and $\gamma_c$, it is possible to arrive at approximations for $f, g$ and $h$ that are useful for the purposes of analytics that can be used to direct meaningful curricular reform efforts. 

Based upon our knowledge of higher education, and the nature of curriculum, it is reasonable to assume that as the complexity of a curriculum increases, a student's ability to complete that curriculum decreases. That is, any complexity measure for a curriculum should relate directly to students' ability to complete the curriculum. If we let $\beta_c$ denote the completion rate of students attempting to complete curriculum $c$, then this assumption can be expressed as:
\[
 \mbox{\rm Assumption 1:} \ \  \uparrow \Psi_c \; \Rightarrow \; \downarrow \beta_c. \ \ \ \ \ \ \ \ \ \ 
\]
We can further decompose this assumption along the structural and instructional complexity components. Specifically, if instructional complexity increases, while keeping the structure of the curriculum unchanged, we would expect completion rates to decrease. This can be expressed as:
\[
 \mbox{\rm Assumption 2:} \ \  \left(\mid \alpha_c, \; \uparrow \gamma_c \right) \; \Rightarrow \; \downarrow \beta_c.
\]
Similarly, we would expect completion rates to decrease if the structure of the curriculum is made more complex~(a notion we will formalize shortly), without changing how courses are taught and delivered. That is,
\[
 \mbox{\rm Assumption 3:} \ \  \left(\uparrow \alpha_c, \; \mid \gamma_c\right) \; \Rightarrow \; \downarrow \beta_c.
\]
For each of these assumptions, we assume the converse assumptions also hold. In particular, 
\[
 \neg\mbox{\rm (Assumption 1)\hc} \ \  \downarrow \Psi_c \; \Rightarrow \; \uparrow \beta_c, \ \ \ \ \ \ \ \ \ \ 
\]
\[
 \neg\mbox{\rm (Assumption 2)\hc} \ \  \left(\mid \alpha_c, \; \downarrow \gamma_c\right) \; \Rightarrow \; \uparrow \beta_c,
\]
and
\[
 \neg\mbox{\rm (Assumption 3)\hc}  \ \  \left(\downarrow \alpha_c, \; \mid  \gamma_c\right) \; \Rightarrow \; \uparrow \beta_c.
\]

Taken together, Assumptions~2 and~3 imply that structural and instructional complexity are independent components. We will make extensive use of this feature as a part of our derivation of useful measures for curricular analytics in this paper. In particular, this feature allows us to fix one of these components, vary the other, and measure the impact on student progression and completion rates. By doing this, we have an indirect means for estimating $\alpha_c$ and $\gamma_c$.  

Next we will consider factors that will be used to quantify the structural and instructional complexities of curricula. The first component we will consider, structural complexity, is easier to quantify due to the fact that it more readily admits a mathematical representation, namely that of a graph. The second component, instructional complexity, is more qualitative in nature, and thus more difficult to characterize mathematically.

%------------------------------------------------
\subsubsection{Structural Factors}\label{struct_factors}
The most natural structural representation for a curriculum is as a directed acyclic graph, constructed by creating a vertex for each course in a curriculum, and by placing a directed edge between two vertices if there is a pre- or co-requisite relationships between the courses the vertices represent. This is in fact the representation we used for the curricula shown Figure~\ref{univ-curric}. More formally, we represent a curriculum $c$ consisting of $n$ courses as a directed graph $G_c = (V,E)$, where each vertex $v_1, \ldots, v_n \in V$ represents a requirement (i.e., course) in $c$, and there is a directed edge $(v_i,v_j) \in E$ from requirement $v_i$ to $v_j$ if $v_i$ must be satisfied prior to the satisfaction of $v_j$. We will refer to $G_c$ as a \emph{curriculum graph}. This graph representation allows us to use graph-theoretic concepts to quantify the structural complexity of curricula. To do this we must relate specific properties of graphs to particular student success factors.

The first thing to note is the critical role prerequisites play in quantifying structural complexity.\footnote{We will use the term prerequisite to refer to both pre- and co-requisites unless explicitly stated otherwise.} The inability of a student to successfully complete a given course in a given term, for whatever reason, is the most direct way to measure a lack of progress within a curriculum. A student may not pass a course in a given term due to not enrolling in the course, enrolling but then withdrawing from the course, or by enrolling but not earning a passing grade. Intuitively, we know that if this course is a prerequisite for other courses in the curriculum, the impact of not passing will be larger than if the course is not a prerequisite for other courses in the curriculum. Thus, the difficulty of completing a curriculum seems to be related to the overall prerequisite structure of its curriculum graph. Furthermore, the prerequisite structures can vary dramatically between two different programs. Indeed, we showed in Figure~\ref{univ-curric} that even the same program at two different schools can have drastically different structures. Our intuition, captured as Assumption 3, is that assuming instructional factors are held constant, student success (as measured by completion rates) and structural complexity are inversely related.  We will provide evidence for this assumption in Section~\ref{practice}. Next we formalize this intuition in order to better characterize structural complexity by considering the factors that impact a student's ability to progress through a curriculum.  

There are numerous graph-related properties that may be used to quantify the structural complexity factors associated with a curriculum. Our contention is that we can express these factors, and the structural complexity of a curriculum as a whole, in terms of the properties of its curriculum graph. An important byproduct of the fact that we can represent curricula mathematically as graphs is that we can rewrite Equation~(\ref{struct_complex}) as:
\begin{equation}
 \alpha_c = h(G_c).
 \label{struct_complex_G}
\end{equation}
That is, the structural properties of a curriculum are completely characterized by its curriculum graph. In the remainder of this paper, we will assume that prerequisite relationships in curriculum graphs are well-formed.  By this we mean there are no directed cycles in $G_c$~(these would make the curriculum impossible to complete) and no forward edges (these are redundant prerequisites the introduce spurious complexity).\footnote{If a curriculum has the prerequisite relationship $v_1 \rightarrow v_2 \rightarrow v_3$, and $v_1$ and $v_2$ are \emph{not} co-requisites, then $v_1 \rightarrow v_3$ would represent an extraneous \emph{forward edge}; $v_1 \rightarrow v_2 \rightarrow v_3$ captures the notion that $v_3$ cannot be taken until $v_1$ has been passed, making $v_1 \rightarrow v_3$ redundant. For information on how to detect forward edges in graphs, see~\cite{CoLeRiSt:09}.} Next, we consider a number of structural factors that can be derived from $G_c$.

\paragraph{Delay and Degrees-of-Freedom Factors.}
Many curricula, particularly those in science, technology engineering and math~(STEM) fields, contain a set of courses that must be completed in sequential order.\footnote{We will use the phrases \emph{pass a course} and \emph{complete a course} synonymously.} For instance, these curricula often contain a math pathway consisting of Calculus I $\rightarrow$ Calculus II $\rightarrow$ Differential Equations $\rightarrow$ Linear Algebra, followed by other major-specific courses that build upon this math background. It is not uncommon in these programs to find prerequisite pathways consisting of seven or eight courses---they span nearly every term in any possible degree plan. The ability to successfully navigate these long pathways without delay is critical for student success and on-time graduation. If any course on the pathway is not completed on time, the student will then be delayed in completing the entire pathway by one term. That is, because of the prerequisite constraints, it is not possible to move any of the courses on the pathway to an earlier term in order to make up time. These prerequisite pathways correspond to paths in $G_c$. A path $p$ in $G_c = (V,E)$, denoted $v_i  \overset{p}{\leadsto} v_j$, $v_i,v_j \in V$, is a sequence of vertices $<v_i, \ldots, v_j>$, all in $V$, such that for each pair of adjacent vertices $v_x$ and $v_y$ in the sequence, if $v_x$ precedes $v_y$ in the sequence, then $(v_x,v_y) \in E$. That is, a path in $G_c$ is simply a set of courses in $c$ that must be taken in sequential order due to prerequisite constraints. We will use $\#(v_i  \overset{p}{\leadsto} v_j)$ to denote the number vertices on path $p$. Note that there may be more than one path between any two pairs of vertices in a curriculum graph $G_c$. 

\begin{figure}
 \centerline{\includegraphics{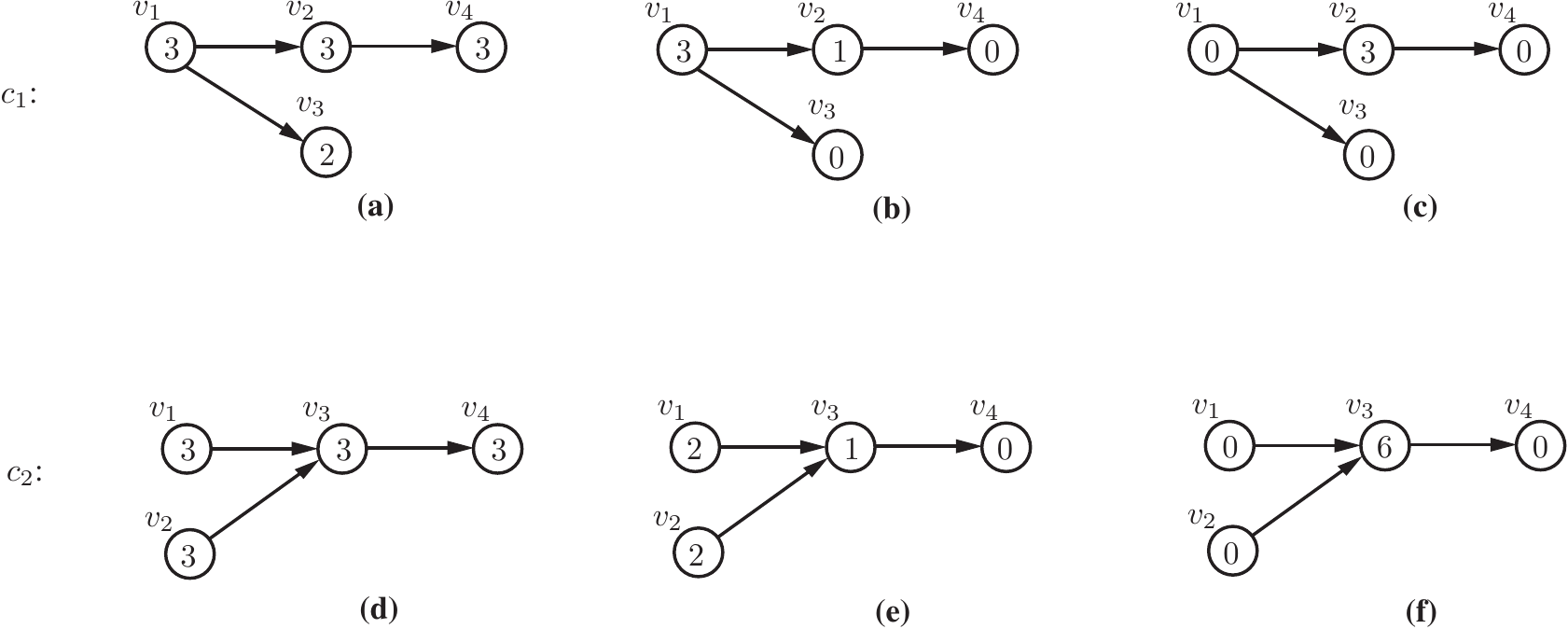}}
 \caption{Two different four-course curricula. Curriculum $c_1$ is shown in {\bf (a)}--{\bf (c)}, and curriculum $c_2$ is shown in {\bf (d)}--{\bf (f)}. Parts {\bf (a)} and {\bf (d)} show the delay factor, parts {\bf (b)} and {\bf (e)} show the blocking factor, and parts {\bf (c)} and {\bf (f)} show the centrality factor associated with each course in these curricula.}
 \label{ex-curric}
\end{figure}

\begin{figure}
 \centerline{\includegraphics[width=6.5in]{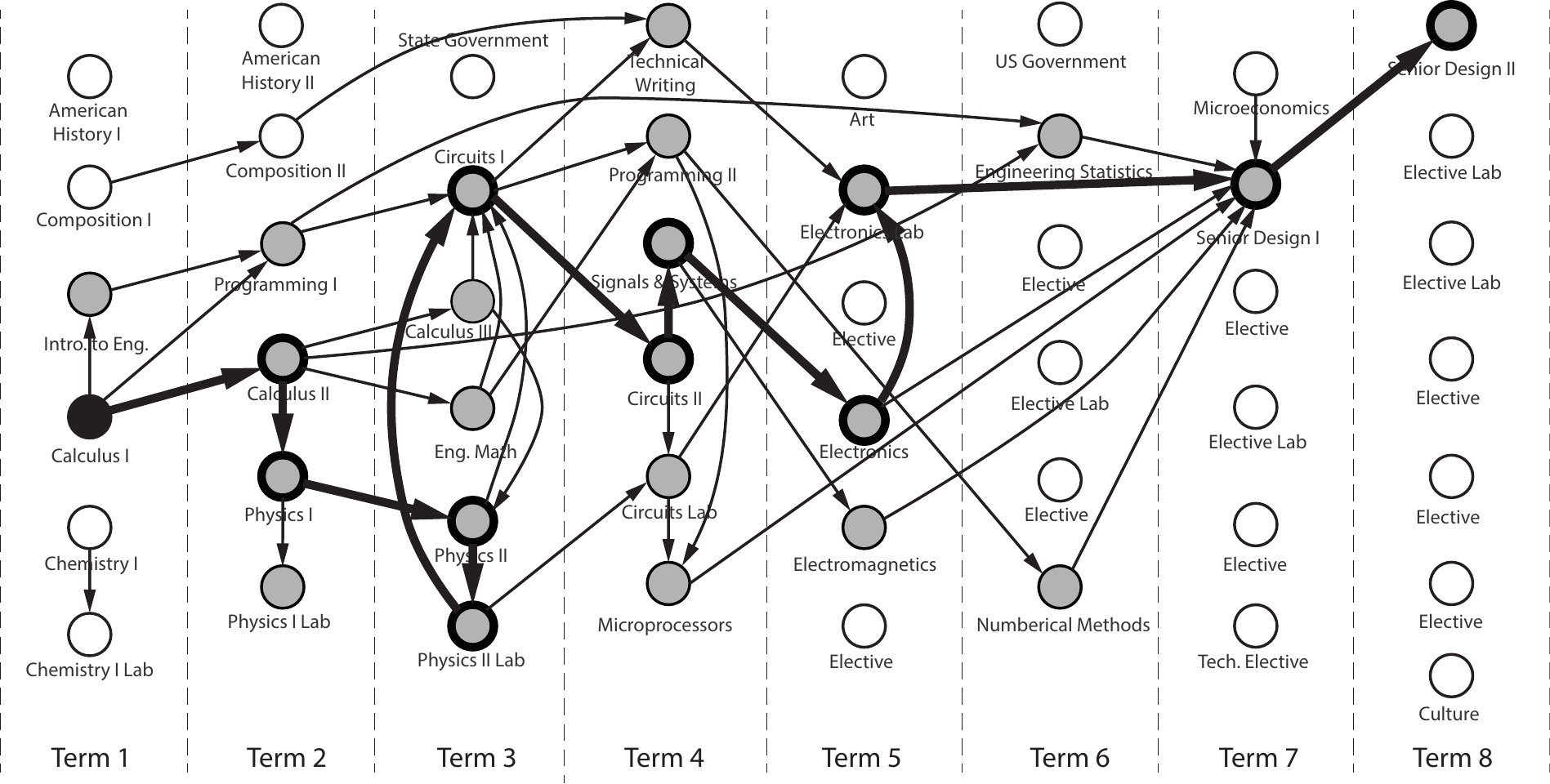}}
 \begin{center}{\bf (a)}\end{center}~\\~\\
 \centerline{\includegraphics[width=6.5in]{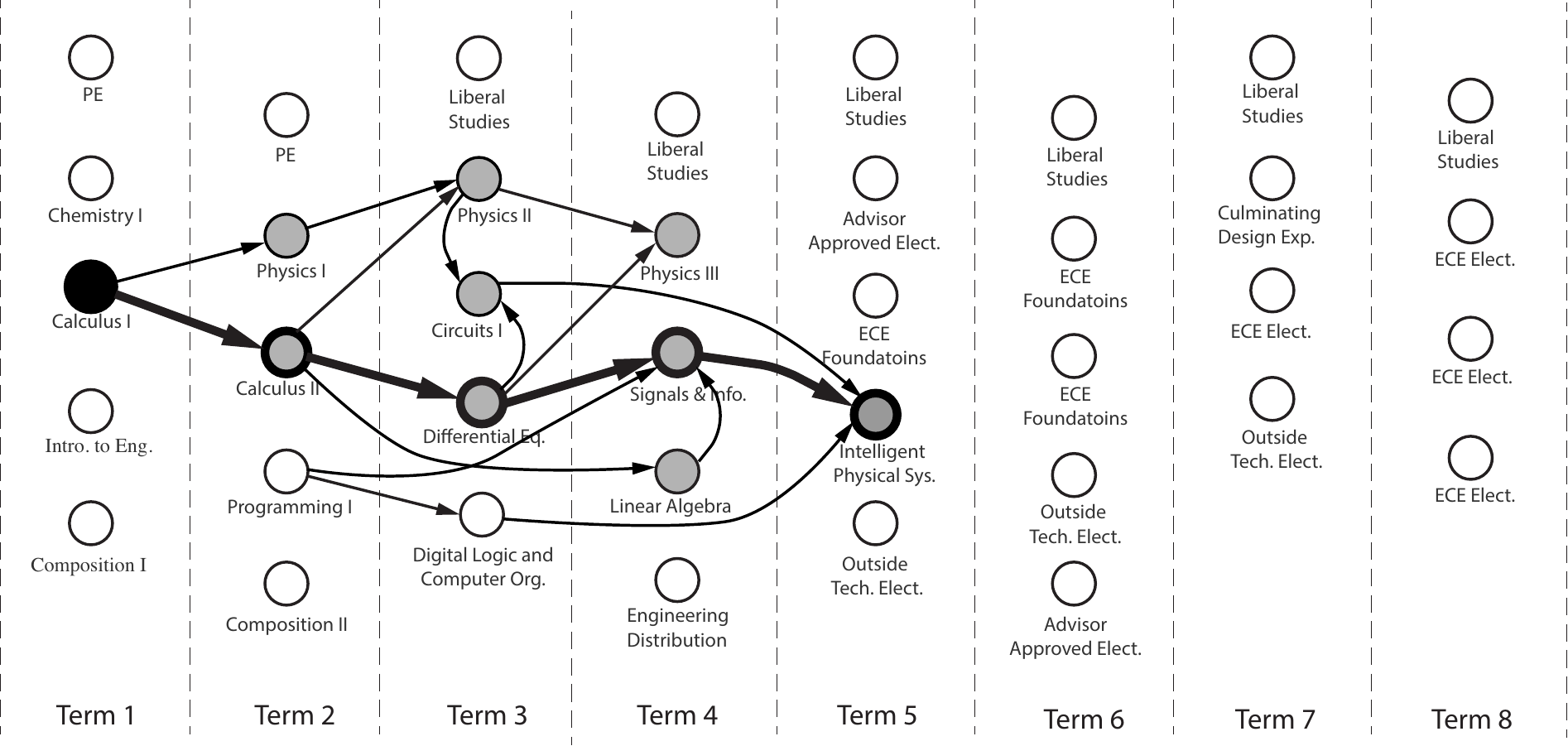}}
 \begin{center}{\bf (b)}\end{center}
 \caption{The longest paths (highlighted edges) and courses blocked (gray vertices) by Calculus I (black vertex) in the two curricula presented in Figure~\ref{univ-curric}. {\bf (a)} The delay factor associated with every course on the longest path is 11 in this curriculum, and the blocking factors associated with Calculus I is 23. {\bf (b)} The delay factor associated with every course on the longest path is 5 in this curriculum, and the blocking factors associated with Calculus I is 9. Note that there are multiple longest paths in each curriculum.}\label{univ-curric-db}
\end{figure}

We define the \emph{delay factor} associated with a given course $v_k$ in  a curriculum $c$, denoted $d_c(v_k)$, as the number of vertices in the longest path in $G_c$ that passes through $v_k$~\citep{Sl:16}. That is,
\begin{equation}
 d_c(v_k) = \max_{i,j,l,m}\left\{\#(v_i  \overset{p_l}{\leadsto} v_k \overset{p_m}{\leadsto} v_j)\right\}.
\end{equation}
In Figures~\ref{ex-curric}~(a) and~(d), we show two simple curricula, $c_1$ and $c_2$, that illustrate the concept of delay.
Each of these curricula have four courses organized over three terms. In curriculum $c_1$ there are two paths to terminal courses, one contains two vertices, and the other has three, while in curriculum $c_2$, both of the longest paths have three vertices. As shown inside of the vertices in Figure~\ref{ex-curric}~(a) and~(d), the delay factor of each vertex is determined by the longest path that it is on. Notice that due to the delay factors, curricula $c_1$ and $c_2$ cannot be completed in fewer than three terms. To get a sense of the delay factors associated with the courses in complete four-year programs, in Figure~\ref{univ-curric-db} we highlight the longest paths in the two curricula from  Figure~\ref{univ-curric}.  
The curriculum in Figure~\ref{univ-curric-db}~(a) has twelve vertices on the longest paths, and the one in Figure~\ref{univ-curric}~(b) has five. The vertices on a longest path in Figure~\ref{univ-curric-db}~(a) span seven of the eight terms in the degree plan. This establishes a lower bound of seven on the number of terms required to complete this curriculum. In Figure~\ref{univ-curric}~(b), the vertices on a longest path span at most five terms, and thus, it is theoretically possible to complete this curriculum in five terms.

We define the delay factor associated with an entire curriculum $c$ as:
\begin{equation}
 d(G_c) = \sum_{v_k \in V} d_c(v_k).
\end{equation}
The delay factor associated with the curriculum shown in Figure~\ref{univ-curric-db}~(a) is 307, while the delay factor associated with the curriculum in Figure~\ref{univ-curric-db}~(b) is only 81.

If a course is standalone---it has no prerequisites, and it is not a prerequisite to another course---it has a delay factor of one, and can be taken in any term.  This freedom to move the course to other terms provides tremendous flexibility if it becomes necessary to reorganize a degree plan in efforts to limit delays as a result of a student not passing a class. We define the \emph{degrees-of-freedom} provided by a curriculum $c$, denoted $z_c$, as the number unique ways in which a curriculum may be reordered on a term-by-term basis over a fixed number of terms, while respecting the prerequisite relationships. For example, allowing for three terms, there is only one way to reorganize the curriculum in Figure~\ref{ex-curric}~(a), course $v_3$ could be moved to the third term, and therefore $z_{c_2} = 2$ over three terms. For the curriculum in Figure~\ref{ex-curric}~(b), there is only one way the curriculum can be organized over three terms, and therefore $z_{c_2} = 1$. This indicates that the latter curriculum is more constrained with regards to how it can be reorganized than the former. The degrees-of-freedom factor is related to the number of weakly connected components in a curriculum graph.\footnote{A \emph{weakly connected component} is a maximal subgraph in a directed graph such that for every pair of vertices $u, v$ in the subgraph, if the edges in the directed graph are treated as if they are undirected, a path exists from $u$ to $v$~\citep{CoLeRiSt:09}.} In summary, as long prerequisite pathways are created within a curriculum, the delay factor associated with the curriculum increases, and the degrees-of-freedom decreases.

\paragraph{Blocking and Reachability Factors.} 
Another structural factor arises when one course serves as the gateway to many other courses in the curriculum. In this case, if a student is unable to pass the gateway course, they are blocked from attempting many of the other courses in the curriculum.  For instance, in many curricula, Calculus I is a foundational first-term course that must be completed before taking other major-specific classes in subsequent terms. It is obvious that a course which is a prerequisite for a large number of other courses in a curriculum is a highly important course in that curriculum.

We will denote the situation where course $v_j$ is reachable from course $v_i$, via any prerequisite pathway, using $v_i \leadsto v_j$, and $v_i \not\leadsto v_j$ will be used if course $v_j$ is not reachable from course $v_i$. The \emph{blocking factor} associated with course $v_i$ in curriculum $G_c = (V,E)$, denoted $b_c(v_i)$, is then given by~\citep{Sl:16}:
\begin{equation}
 b_c(v_i) = \sum_{v_j \in V} I(v_i,v_j)
\end{equation}
where $I$ is the indicator function:
\begin{equation}
 = \begin{dcases}
                       1, & \mbox{if $v_i \leadsto v_j$;} \\
                       0, & \mbox{if $v_i \not\leadsto v_j$}.
                  \end{dcases} 
                  \label{path_indicator}
\end{equation}
In Figures~\ref{ex-curric}~(b) and~(e) we show the blocking factors associated with the two four-course curricula $c_1$ and $c_2$. Notice that course $v_1$ in curriculum $c_1$ has the largest blocking factor. In Figure~\ref{univ-curric-db} we have shaded the vertices that are blocked by the Calculus~I course in each of the curricula from Figure~\ref{univ-curric}. Specifically, the blocking factor of Calculus~I in the curricula in Figure~\ref{univ-curric-db}~(a) is 23, which corresponds to half of the courses in that curriculum. The blocking factor of Calculus~I in the curriculum in Figure~\ref{univ-curric-db}~(b) is 9, which corresponds to a quarter of the courses in that curriculum.

The blocking factor associated with an entire curriculum $c$ is defined as the sum of blocking factors of all vertices in $G_c$:
\begin{equation}
 b(G_c) = \sum_{v_i \in V} b_c(v_i).
\end{equation}
For instance, the blocking factor associated with the curriculum in Figure~\ref{univ-curric-db}~(a) is 226, while  the blocking factor associated with the curriculum in Figure~\ref{univ-curric-db}~(b) is 30.

We may define a corresponding \emph{reachability factor} as the number of courses that must be completed before one would be allowed to take a given course. That is, the \emph{reachability factor} associated with course $v_i$ in curriculum $G_c = (V,E)$, denoted $r_c(v_i)$, is  given by:
\begin{equation}
 r_c(v_i) = \sum_{v_j \in V} I(v_j,v_i),
\end{equation}
where $I$ is the indicator function defined in Equation~(\ref{path_indicator}). 
The reachability factor associated with curriculum $c$ is:
\begin{equation}
 r(G_c) = \sum_{v_i \in V} r_c(v_i).
\end{equation}
Note that  $b(G_c)$ can also be computed as the sum of the out-degrees of all vertices in $G_c$, and that $r(G_c)$ can be computed by summing the in-degrees of all vertices in $G_c$. Because the total in-degree in a directed graph must equal the total out-degree, it follows that $b(G_c) = r(G_c)$.

\paragraph{Centrality Factor.}
Finally, we consider a factor that is based on the notion of centrality. We can define a course as being central to a curriculum if it requires a number of foundational courses as prerequisites, and the course itself serves as a prerequisite to many additional discipline-specific courses in the curriculum, typically in the junior and senior years. For instance, in electrical engineering programs, Circuits I is central (see Figure~\ref{univ-curric}); in accounting, Principles of Financial Accounting is central; in mechanical engineering, Mechanics (statics and dynamics) is central; and in chemistry, Organic Chemistry is central. 

In the study of networks, numerous notions of centrality have been defined in order to address particular applications~\citep{Ne:10}.
Perhaps the closest to the idea of determining central courses in a curriculum is provided by the \emph{betweenness centrality} measure. The betweenness centrality of vertex $v \in V$ in a graph $G = (V,E)$ is proportional to the number of shortest~(geodesic) paths containing $v$ between all pairs of vertices in $V$~\citep{Fr:77}. Nodes in a communication network that have high betweenness centrality are important because routing algorithms are designed so that much of the network traffic passes through these nodes. In social networks, betweenness centrality is used to identify the most influential people in a group of people that communicate with one another.

In curricula graphs we are interested in assessing how essential knowledge~(learning outcomes) attained in one course is used in subsequent courses. That is, we are interested in the flow of knowledge between the network of courses in a curriculum, and in identifying those courses that are most central in this flow. In the case of curricula graphs, however, the flow of knowledge through \emph{all} paths in the network is important, not just the flow through shortest paths.  Furthermore, as we have already discussed when considering the delay factor, it is actually the flow of knowledge through the longest paths that matter in curricula. Thus, we define the centrality of a course in a curriculum according to the number of long paths that include the course. More formally, consider a curriculum graph $G_c = (V,E)$, and a vertex $v_i \in V$. Furthermore, consider all paths between every pair of vertices $v_j, v_k \in V$ that satisfy the following conditions:
\begin{enumerate}
 \item $v_i, v_j, v_k$ are distinct, i.e., $v_i \neq v_j, v_i \neq v_k$ and $v_j \neq v_k$;
 \item there is a path from $v_j$ to $v_k$ that includes $v_i$, i.e., $v_j \leadsto v_i \leadsto v_k$;
 \item $v_j$ has in-degree zero, i.e., $v_j$ is a ``source'' node; and
 \item $v_k$ has out-degree zero, i.e., $v_k$ is a ``sink'' node.
\end{enumerate} 
Let $P_{v_i} = \{p_1, p_2, \ldots\}$ denote the set of all paths that satisfy these conditions. Then the centrality of $v_i$, denoted $q(v_i)$, is defined as
\begin{equation}
 q(v_i) = \sum_{l=1}^{\left| P_{v_i} \right|} \#(p_l).
\end{equation}
According to this definition, only vertices on paths containing at least three vertices can have a centrality measure greater than zero, and all source and sink vertices will have a centrality measure of zero. 

Figures~\ref{ex-curric}~(c) and~(f) show the centrality factors associated with the courses in curricula $c_1$ and $c_2$. The most central course in either curriculum is course $v_2$ in curriculum $c_2$. In Figure~\ref{univ-curric-db}~(a), Circuits I has a centrality factor of 15, while in Figure~\ref{univ-curric-db}~(b) the same course has a centrality factor of 516.

The aforementioned factors represent quantities that we believe capture important structural features that influence student progression through a curriculum, and we will demonstrate a number of ways to use them in Section~\ref{practice}; however, we do not claim these are the only useful metrics that can be derived from the properties of curriculum graphs. Before considering how to use these structural factors, let us first consider factors associated with instructional complexity.

\subsubsection{Instructional Factors}
As we have mentioned, the factors associated with instructional complexity tend to be qualitative in nature. For instance, it is difficult to arrive at a precise measure for the quality of instruction that directly relates to student progression. Similarly, it is hard to precisely quantify the impact student support programs have on student progression within particular curricula. We can think of these instructional factors as hidden variables that cannot be directly measured, and our challenge then becomes one of imputing instructional complexity from observable data. Recall that our goal, expressed as Equation~(\ref{curric_complex}), is to characterize curricular complexity in terms of its structural and instructional factors. According to Assumption~1, curricular complexity varies inversely with completion rate.  Assumption~2 further stipulates that if the structural complexity of a curriculum is held constant, the completion rate will change in inverse proportion to changes in instructional complexity. Thus, we seek a measure that relates instructional factors to student progression within a curriculum. The grades earned by students taking the courses in a curriculum provide such a measure. Specifically, a non-passing grade earned by a student in any course within a curriculum, by definition, means that this student cannot immediately progress beyond that point in the curriculum. Because of this direct relationship between course pass/fail rates and curricular progression, we will use course pass/fail rates as a proxy for instructional complexity in the experiments described in Secion~\ref{practice}.

It is worth mentioning that the specific scores earned in courses contain more information related to instructional complexity than pass/fail rates alone. For instance, we expect that on average, students earning higher grades will progress more rapidly through a curriculum than those earning lower grades. A higher grade in a course  is an indicator that a student has attained the learning outcomes associated with the course to a greater extent than those earning lower grades. This advantage should then propagate to other courses in the curriculum, particularly those that are highly reliant on these learning outcomes.

In summary, rather than attempting to disaggregate and measure the various factors that influence instructional complexity, a difficult task given the qualitative nature of these factors, we will use course grades, and in particular pass/fail rates, as an indirect measure of instructional complexity.  For the purposes of the experiments described in the next section, where the focus is on quantifying structural complexity, this will suffice.

%------------------------------------------------
\section{Curricular Analytics---Practice}\label{practice}
In this section we demonstrate how the framework described in Section~\ref{theory} supports methodologies that can be used to derive useful analytics around curriculum. Specifically, we will describe methodologies that have allowed us to gain a better understanding of curricular complexity by exploring the interplay between its main components, structural complexity and instructional complexity. In particular, we will show how to use Equation~(\ref{curric_complex}) and Assumptions~1--3 to ascertain important properties related to structural and instructional complexities. The exact relationship between these two components is difficult to characterize; however, an important property of the assumptions we have made around the decomposition of curricular complexity is that we do not need exact measures of one component in order to uncover important properties of the other. A factor that complicates this approach is that it is difficult to conduct \emph{in situ} experiments on curriculum. The main complication being that this experimentation may impact the success or failure of actual students. More specifically, because the flow of actual students through a curriculum, and therefore student completion rates, would be impacted by any controlled experiments that involved varying either structural or instructional complexity in ``live'' curricula, we developed a simulation model that allows us to study the impact of these components on hypothetical students within a simulated environment.

\subsection{Structural Complexity}
In order to better understand structural complexity we constructed an experiment that isolated the impact of curricular structure on progression.  In particular, we used the experimental design described in this section to obtain a better understanding of the graph properties that characterize structural complexity. The experiment was constructed to make use of Assumption~3 in order to indirectly explore the impact of structural complexity. Taken together, Assumption~3 and its converse imply that if we fix instructional complexity and vary structural complexity, then curriculum completion rates should vary indirectly in proportion to changes in structural complexity. For the purposes of this experiment, we used the aforementioned proxy for instructional complexity, namely the pass/fail rates of the courses within a curriculum. Based upon the discussion provided in the previous section, we assert that it is reasonable to assume that instructional complexity varies directly in proportion to the pass/fail rates of the courses within a curriculum. That is,
\[
 \mbox{\rm Assumption 4:} \ \  \gamma_c \sim \mbox{\rm  course pass/fail rates}.
\]
This assumption can be used to construct experiments around Assumptions~2 and~3.

\begin{figure}
 \centerline{\includegraphics{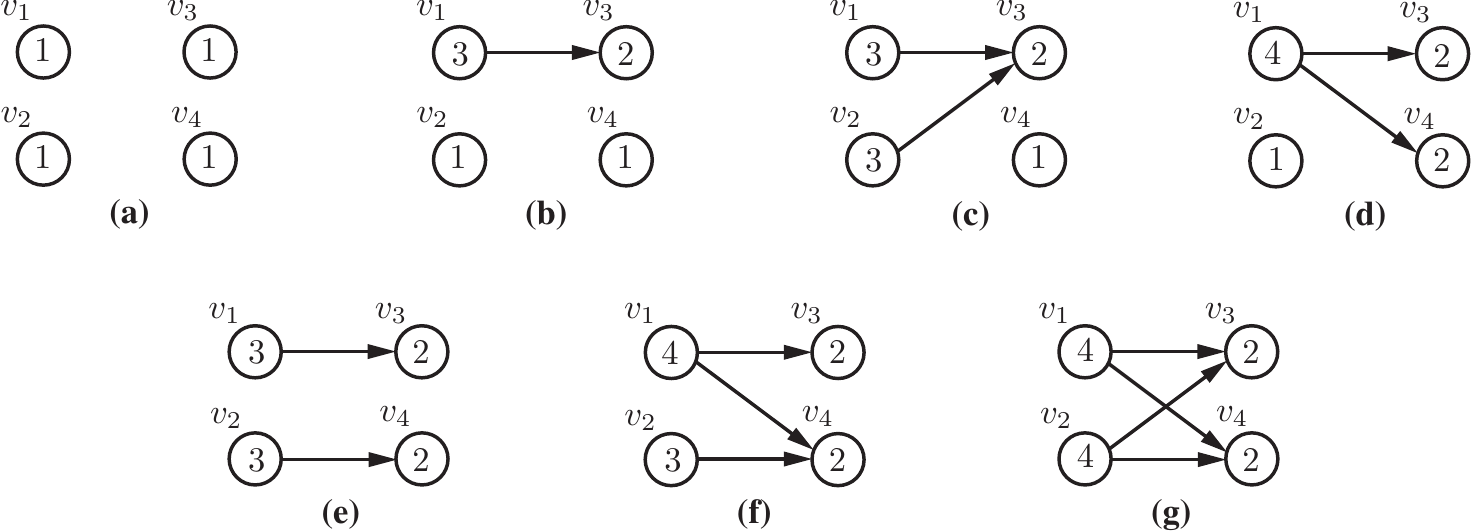}}
 \caption{All possible balanced two-term curricula consisting of four courses (excluding co-requisites), in order of increasing structural complexity.  Courses $v_1$ and $v_2$ are scheduled for term 1, and courses $v_3$ and $v_4$ are scheduled for term 2. The structural complexity of each course is shown inside of its corresponding vertex.  In order, from curricula {\bf (a)} to {\bf (g)}, the structural complexities of the curricula using Equation~(\ref{struct_complex_fit}) are: 4, 7, 9, 9, 10, 11 and 12.}\label{4-course}
\end{figure}

The experimental design described here involves a series of trials that make use of Assumptions 3 and 4 in order to characterize function $h$; each trial can be expressed compactly as follows:
\begin{align}
 \big( \overbrace{\alpha_c}^{\mbox{\rm\footnotesize vary}}, \; \overbrace{\mbox{\rm course pass/fail rates}}^{\mbox{\rm\footnotesize fix}} \big) \ \sim \overbrace{\beta_c}^{\mbox{\rm\footnotesize measure}} \label{exp_struct} \\
 { \mbox{\rotatebox[origin=r]{-15}{$\xleftarrow{\makebox[1.55cm]{}}$}} {\mbox{\rm\footnotesize correlate}} \mbox{\rotatebox{15}{$\xrightarrow{\makebox[1.75cm]{}}$}} \ \ \ \ \ \ \ \ \ \ \nonumber }
\end{align}
A particular trial of this experiment considers the set of all curricula consisting of a fixed number of courses offered over a given number of terms~(e.g., four courses offered over two terms). All of the curricula in a set contain the same courses, with fixed pass/fail rates. Thus, the curricula in a set only differ from one another in structure. For instance, in Figure~\ref{4-course} we show all balanced four-course curricula that can be constructed over two terms.\footnote{In a \emph{balanced} curriculum, courses are  equally distributed over terms.}
Each run within a trial involves simulating student flow through a curriculum in order to compute the average completion rate, i.e., graduation rate. The completion rates among the curricula in a set are then compared to the structural factors mentioned in Section~\ref{struct_factors} in order to determine the curriculum graph properties that are most highly correlated with completion rates.

A number of variables in these simulations impact curriculum completion rates. Important variables include, the course pass/fail rates, the number of terms allowed to complete the curriculum, the maximum number of courses that may be attempted in a term, and the behavior of students who fail to complete a course. For each trial, we assumed the same pass rate for all courses in a curriculum, and all other variables were also held constant. The only thing that varied within a given trial was the structure of the curriculum.

With regards to course-taking behavior, in each trial we allowed students to attempt one more course per term (if possible) than is specified in the degree plan. If a student failed to complete a course in a particular term, they attempted to complete it again in the next term; that is, students were not allowed to ``stop out.'' Note that this is the most optimistic assumption possible with regards to course re-taking---it corresponds to entirely resilient students who will continue to enroll in a course no matter how many times they fail it. Thus, the graduation rates these trials yield will be higher than those of actual students who have the option of stopping out. Keep in mind, however, that this is not the behavior we are attempting to model. These trials are aimed at relating a student's likelihood of completing a curriculum to the structural complexity of that curriculum. As long as the stop-out behavior is consistent across all curricula in the experiment, the relationship will be valid. We hypothesize that there may actually be a slight bias in our stop-out assumption. The very nature of high structural complexity curricula is that they slow progression, thereby presenting students with more opportunities to stop out prior to completion. Recognize, however, that in the case of the aforementioned experiment, this assumption will yield conservative results. That is, the structural complexity of a curriculum is likely to have a larger negative impact on student progression than these experiments predict.

\begin{figure}
 \centerline{\includegraphics[width=5in]{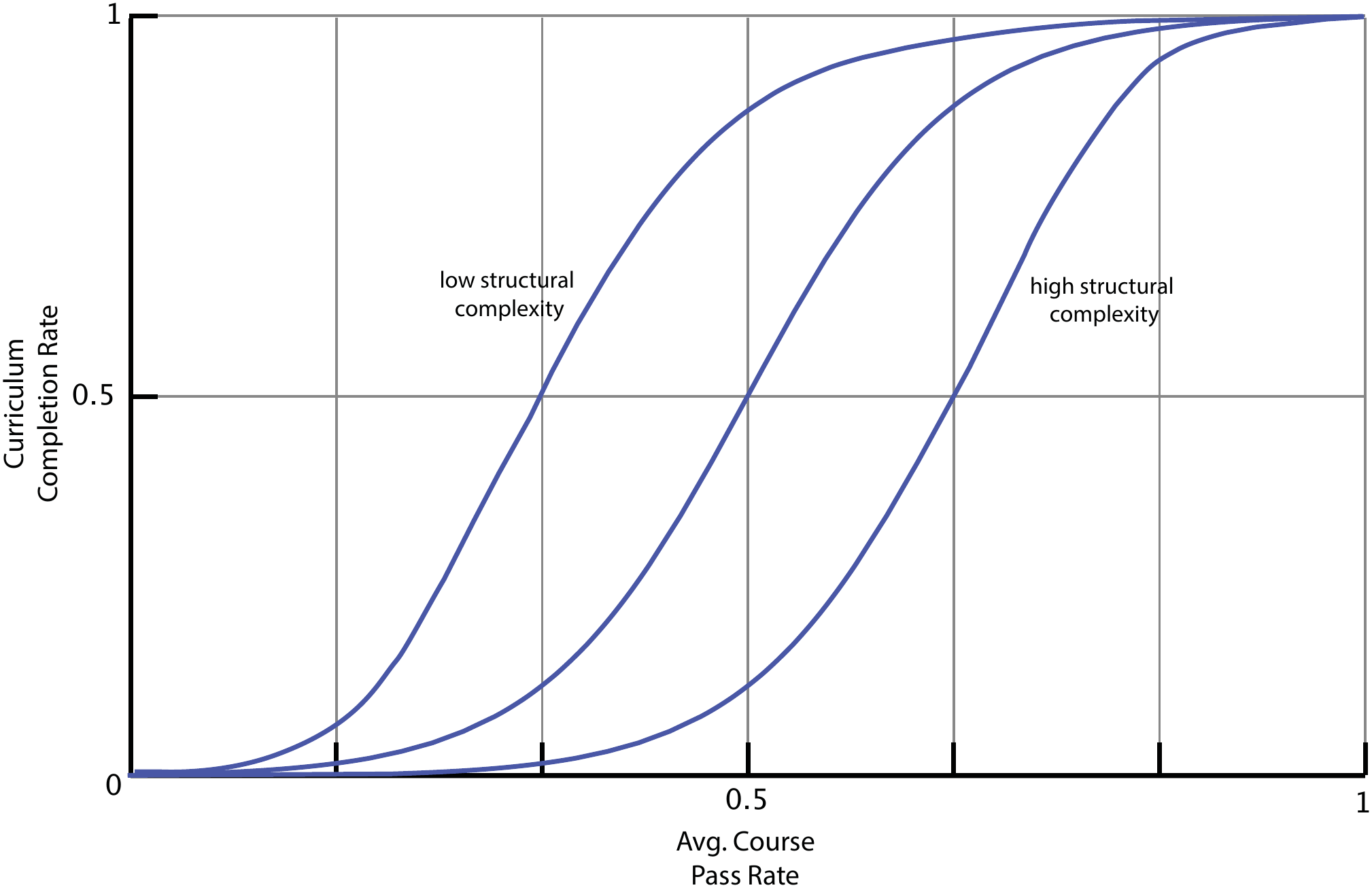}}
 \caption{Curriculum completion rates versus average pass rates (a proxy for instructional complexity) for three curricula with increasing structural complexity.}\label{logistic}
\end{figure}

Finally, let us consider the matter of how to set course pass/fail rates. For a given curriculum, there are two extremes. First there is the case where the instructional complexity is such that every student completes every course in the curriculum on the first attempt, i.e., no student ever fails to complete a class.  In this case, the structural complexity of a curriculum is irrelevant---the on-time graduation rate will be 100\% no matter the structure of the curriculum. At the other extreme, the instructional complexity is such that no student is ever able to complete a course. In this case, student progress is again independent of the structural complexity, the on-time graduation rate is 0\% for any curriculum graph. For any situation inbetween these two extremes, it is the combination of the structural and instructional complexities that determines student progress.  Figure~\ref{logistic} shows completion rate curves for curricula with three different structural complexities.
This figure notionally captures the behavior we observed, namely that there is a large range over which the interplay between these two quantities is roughly linear. Furthermore, as more time is allowed for completion, the individual curves shift towards the right.  

Table~\ref{Ex-gradrate} shows the simulated graduation rates for each of the curricula in Figure~\ref{4-course} when all course pass rates are set at 50\%. 
\begin{table}
 \begin{minipage}{0.5\textwidth}
 \centering
 \begin{tabular} { c | c c c c}
               & $t_1$    &      $t_2$      &      $t_3$       &      $t_4$  \\ \hline 
  $v_1$   &  50       &       75        &         87.5        &       93.75   \\ 
  $v_2$   &  50       &       75        &         87.5        &       93.75   \\ 
  $v_3$   &  50       &       75        &         87.5        &       93.75   \\ 
  $v_4$   &    0        &      50        &         75        &       87.75   \\ \hline
  grad. rate & 0        &   28.13     &        50.24     &      {\bf 72.30} \\ \hline
 \end{tabular}
 \begin{center}
   {\bf (a)}
 \end{center}
 \end{minipage}
  \begin{minipage}{0.5\textwidth}
 \centering
  \begin{tabular} { c | c c c c}
               & $t_1$    &      $t_2$      &      $t_3$       &      $t_4$  \\ \hline 
  $v_1$   &  50       &       75        &         87.75        &       93.75   \\ 
  $v_2$   &  50       &       75        &         87.75        &       93.75   \\ 
  $v_3$   &  0        &       25         &          50            &       68.75   \\ 
  $v_4$   &   50        &      75        &         87.75      &        93.75   \\ \hline
  grad. rate & 0        &   10.55      &         33.78      &         {\bf  56.65} \\ \hline
 \end{tabular}
  \begin{center}
   {\bf (b)}
 \end{center}
 \end{minipage}~\\~\\

 \begin{minipage}{0.5\textwidth}
 \centering
  \begin{tabular} { c | c c c c}
               & $t_1$    &      $t_2$      &      $t_3$       &      $t_4$  \\ \hline 
  $v_1$   &  50       &       75        &         87.75        &       93.75   \\ 
  $v_2$   &  50       &       75        &         87.75        &       93.75   \\ 
  $v_3$   &    0       &       12.5      &        34.38        &       55.47   \\ 
  $v_4$   &  50        &      75        &         87.75        &       93.75   \\ \hline
  grad. rate & 0        &     5.27      &         23.23      &        {\bf 45.71}  \\ \hline
 \end{tabular}
  \begin{center}
   {\bf (c)}
 \end{center} 
  \end{minipage}
 \begin{minipage}{0.5\textwidth}
 \centering 
 \begin{tabular} { c | c c c c}
               & $t_1$    &      $t_2$      &      $t_3$       &      $t_4$  \\ \hline 
  $v_1$   &  50       &       75        &         87.75        &       93.75   \\ 
  $v_2$   &  50       &       75        &         87.75        &       93.75   \\ 
  $v_3$   &    0       &       25        &         50           &         68.75   \\ 
  $v_4$   &    0        &      25        &         50           &         68.75   \\ \hline
  grad. rate & 0        &     3.52      &       19.25        &         {\bf 41.54}  \\ \hline
 \end{tabular}
 \begin{center}
   {\bf (d)}
 \end{center} \end{minipage}~\\~\\
 
 \begin{minipage}{0.5\textwidth}
 \centering 
 \begin{tabular} { c | c c c c}
               & $t_1$    &      $t_2$      &      $t_3$       &      $t_4$  \\ \hline 
  $v_1$   &  50       &       75        &         87.75        &       93.75   \\ 
  $v_2$   &  50       &       75        &         87.75        &       93.75   \\ 
  $v_3$   &    0       &       25        &         50            &        68.75   \\ 
  $v_4$   &    0        &      25        &         50            &        68.75   \\ \hline
  grad. rate & 0        &     3.52     &         19.25      &         {\bf 41.54}  \\ \hline
 \end{tabular}
 \begin{center}
   {\bf (e)}
 \end{center}
 \end{minipage}
 \begin{minipage}{0.5\textwidth}
 \centering 
 \begin{tabular} { c | c c c c}
               & $t_1$    &      $t_2$      &      $t_3$       &      $t_4$  \\ \hline 
  $v_1$   &  50       &       75        &         87.75        &     93.75   \\ 
  $v_2$   &  50       &       75        &         87.75        &       93.75   \\ 
  $v_3$   &    0       &       25        &         50             &       68.75   \\ 
  $v_4$   &    0        &    12.5        &        34.38        &      55.47   \\ \hline
  grad. rate & 0        &      1.76     &         13.24       &       {\bf 33.52}  \\ \hline
 \end{tabular}
  \begin{center}
   {\bf (f)}
 \end{center}
 \end{minipage}~\\~\\

 \centering 
 \begin{tabular} { c | c c c c}
               & $t_1$    &      $t_2$      &      $t_3$       &      $t_4$  \\ \hline 
  $v_1$   &  50       &       75        &         87.75        &       93.75   \\ 
  $v_2$   &  50       &       75        &         87.75        &       93.75   \\ 
  $v_3$   &    0       &       12.5      &        34.38        &       55.47   \\ 
  $v_4$   &    0        &      12.5      &        34.38        &       55.47   \\ \hline
  grad. rate & 0        &       1.58    &         9.10         &        {\bf 27.04}  \\ \hline
 \end{tabular}
 \begin{center}
   {\bf (g)}
 \end{center}
 \caption{The graduation rates associated with the curricula provided in Figure~\ref{4-course} under a simple set of instructional complexity assumptions. The entry associated with $v_i$ and $t_j$ in a given table is the percentage of students who have passed course $v_i$ after term $t_j$. The structural complexities of the curricula shown in Figure~\ref{4-course} are monotonically increasing from curriculum (a) through (g). The 200\% graduation rate for these same curricula, shown in bold type face in Tables~(a) through (g) above, decreases monotonically from curriculum (a) through (g).} \label{Ex-gradrate}
\end{table}
The tables in this figure show graduation rates for students at 100\% (i.e., on time), 150\% and 200\% time. Notice that as we allow more time, the graduation rates in each curriculum increase, as we would expect. In addition, as we move from curriculum~(a) to~(g), the 100\%, 150\% and 200\% graduation rates all decrease monotonically. In other words, the behavior we would expect to encounter \emph{in situ} is monotonically decreasing completion rates as students attempt to progress through curricula~(a) through~(g). Our goal then is to find a collection of structural factors that are highly correlated to the behavior shown in Table~\ref{Ex-gradrate}.

\begin{figure}
 \centerline{\includegraphics[width=5in]{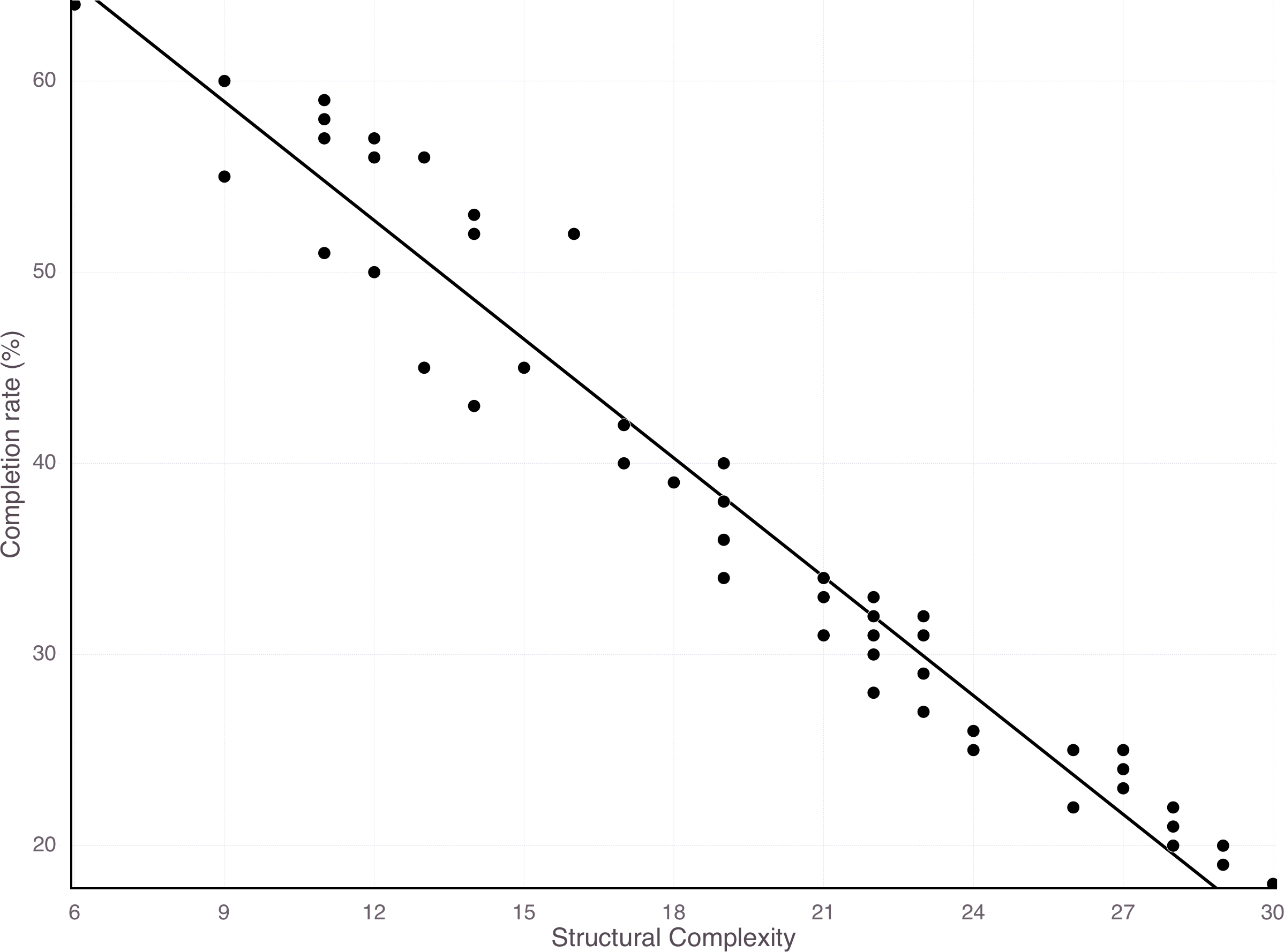}}
 \caption{The dots in this figure correspond to the simulated completion rates, allowing five terms to complete, of all six-course curricula balanced over three terms. The line, give by Equation~(\ref{struct_complex_fit}), is the least squares fit provided by linear regression with a coefficient of determination, $r^2$, value of $0.955$.}\label{linear_reg}
\end{figure}

The experiments we constructed, summarized in Equation~(\ref{exp_struct}), involved first simulating the completion rates in all balanced curricula of a given size using the same pass rate for all courses. Next we enumerated the various linear combinations of the structural complexity factors described in Section~\ref{struct_factors}. Finally, we used linear regression to correlate the different linear combinations of complexity factors to curriculum completion rates. The curricula we considered included all four course curricula balanced over two terms, and all six course curricula balanced over three terms.  A good fit over a wide range of course pass rates was provided by the following function:
\begin{align}
 h(G_c) & = d(G_c) + b(G_c) \nonumber \\
             & = \sum_{v_k \in V} \Big(d_c(v_k) + b_c(v_k)\Big).\label{struct_complex_fit}
\end{align}
In Figure~\ref{linear_reg} we show the fit that is obtained by using this linear combination to model the structural complexity of all six course curricula balanced over three terms.

It is interesting to note that only two structural factors, delay and blocking, are required to provide a good characterization of structural complexity. The reason for this is that the three other structural factors we defined---reachability, degrees-of-freedom and centrality---are all linearly dependent on the delay and blocking factors, and therefore with regards to computing structural complexity, they provide redundant information. This does not mean that these three factors are unimportant. On the contrary, we believe they are helpful in providing insights about curricula, and that they will prove useful in the development of other curriculum-related metrics. 

The importance of Equation~(\ref{struct_complex_fit}) is that it provides a unitless measure for structural complexity that can be applied to any curriculum, and it is important to keep in mind that the experimental design explicitly relates this measure to the likelihood that a student can complete a curriculum.  Notice also, that Equation~(\ref{struct_complex_fit}) provides a vertex-by-vertex contribution to the overall structural complexity of a curriculum. That is, course $v_k$ has structural complexity 
$d_c(v_k) + b_c(v_k)$, and therefore we have a means of characterizing the structural complexity of a curriculum as well as the individual courses in the curriculum. Thus, from a structural complexity perspective, we can identify the most important courses in a curriculum, the most complex term, etc.  Inside of each of vertex in Figure~\ref{4-course}, we show the value of this course structural complexity measure.  Notice that the structural complexities of the curricula, computed using Equation~(\ref{struct_complex_fit}) increase monotonically from curriculum~(a) through~(g). 

%------------------------------------------------
\section{Curricular Analytics---Application}\label{application}
In this section we provide a few examples of how the methodologies developed in Section~\ref{practice} can be used to analyze curricula and guide reform efforts aimed a improving student success outcomes. In many cases, curricular reforms are motivated by faculty opinion supported with anecdotal evidence. This is not meant to fault faculty who endeavor to enhance their programs by bringing their experiences to bear on curriculum redesign discussions. Rather, it is indicative of the lack of formal frameworks that can be applied to the study of curricula. Historically, reform efforts have basically treated a curriculum as a black box system, with student characteristics, including prior preparation, as inputs, and student success rates as outputs. A curricular reform in this case was viewed as a change to the internal structure of the black box, and the efficacy of the reform was judged by the impact on student outcomes, which often take years to materialize. 

The contribution of this work is that it provides a means to quantify the curricular system within the black box---in essence, curricular analytics serves to reduce system opacity. In doing so, it provides a means for directly analyzing the curricular system itself, thereby supporting predictive analytics as an integral part of of curricular redesign efforts. Below we provide specific examples of the usefulness of this approach. 

\subsection{Comparing Curricula}\label{compare}
A straightforward application of curricular analytics involves using them to compare the complexities of academic programs. Specifically, given a collection of curricula $C = \{c_1, \ldots, c_n\}$, one may be interested in comparing and contrasting the complexities of the curricula within $C$, or there may be interest in comparing the structural complexities of the curricula in $C$ to actual student success outcomes. As an example of the former, consider the case when $C$ is comprised of the set of all curricula provided in a given field of study, as classified by CIP code (NCES, 2018).\nocite{CIP:18} Quantifying the disparities in the perceived complexities of similar programs at different schools, and the impact of these disparities on student success, was in fact the major impetus for the research described in this paper. The comparison of the two curricula shown in Figure~\ref{univ-curric} is just one example of the significant structural variations we found between similar programs at different institutions. Not surprisingly, we have observed that the structural variances among curricula in $C$ is greater in those fields that tend to have highly correlated sequential knowledge development as one progresses from novice to expert in the field. This is most prevalent in STEM fields.

Given that we can use the curricular complexity metrics described in this paper to order the elements of $C$ according to their structural complexity, the next consideration is how we might use this information. We have observed curriculum reformers use this information in a number of different ways. First, reformers have used this ordering to compare their program to programs at other institutions, as a means of investigating the various curriculum reform options that might be possible. The set $C$ might also represent the historical set of curricula offered by a single program over the years. Such a collection provides useful comparison benchmarks that program faculty may use as they contemplate additional curricular modifications. For example, these benchmarks can be used to evaluate the extent to which particular modifications will impact structural complexity, and therefore expected student completion rates. Furthermore, if historical student success data has been collected by the program, it can be used to more accurately calibrate the expected results of reform efforts. Similarly, faculty can use a given benchmark curriculum to estimate the impact that instructional reform might have on that curriculum. For instance, the simulation model we previously described could be used to evaluate the impact that improving a given course's pass rate would have on overall completion rates. This, in turn, could be leveraged to a curriculum-wide sensitivity analysis that could be used to estimate the best application of limited resources toward individual course improvement efforts.

Next, consider the case where $C$ consists of the curricula associated with all of the programs offered by a particular institution. At the University of New Mexico we ranked these curricula according to their structural complexities, and we then correlated them to the actual six-year graduation rates of actual students using linear regression. The resulting model indicated that every 17 point decrease in structural complexity corresponds to a 1\% increase in the six-year graduation rate. This observation served as an impetus for many programs to reduce the structural complexities of their curricula.

This leads logically to the question of how structural complexity relates to program quality. Our experience is that many assume program quality increases with increasing structural complexity; that is, that complexity equals quality. Our initial investigation of this relationship, however, reveals quite the opposite. Specifically, we have observed an inverse correlation between the perceived quality of engineering programs~(as determined by US News \& World Reports ranking) and the structural complexities of these programs. The nature of this relationship merits further investigation. We conjecture, however, that an Occam's razor-like principle applies to curricula. Namely, the simplest curricula~(in terms of structural complexity) that allows students to attain a program's learning outcomes yields the best student success outcomes and therefore the highest quality program.

\subsection{Curricular Design Patterns}\label{design_patterns}
Another interesting application of curricular analytics involves using them to analyze important design patterns contained within curricula. \cite{AlIsSi:77} first articulated the notion of  \emph{design patterns} as a means of capturing, in general terms, solutions to particular architectural design challenges; that is, the design challenges confronted by architects. \cite{BeCu:87} subsequently extended this idea by  recognizing that software design patterns can aid in the development of complex software systems by providing developers with a reusable set of solutions to common software design problems.  The overarching notion is that design patterns document useful solutions to design challenges that may occur in many different contexts within a given problem domain. In doing so, the design patterns themselves become a vocabulary for designers to use when discussing particular designs.

\cite{HeHiSlAb:17} describe the application of design patterns within the curricular problem domain. They define a \emph{curricular design pattern} as a collection of curricular and co-curricular learning activities intentionally structured so as to allow students to attain a set of learning outcomes within a given educational context. The typical manner in which learning outcomes are attained is through the offering of courses that have have pre- and co-requisite relationship between them. That is, learning activities are structured as courses that must be passed in sequence in order to attain the learning outcomes. The general notion is that prerequisite courses contain learning activities that must be successfully completed in order to attain the learning activities that occur in follow-on courses. 

\begin{figure}
  \centerline{\includegraphics{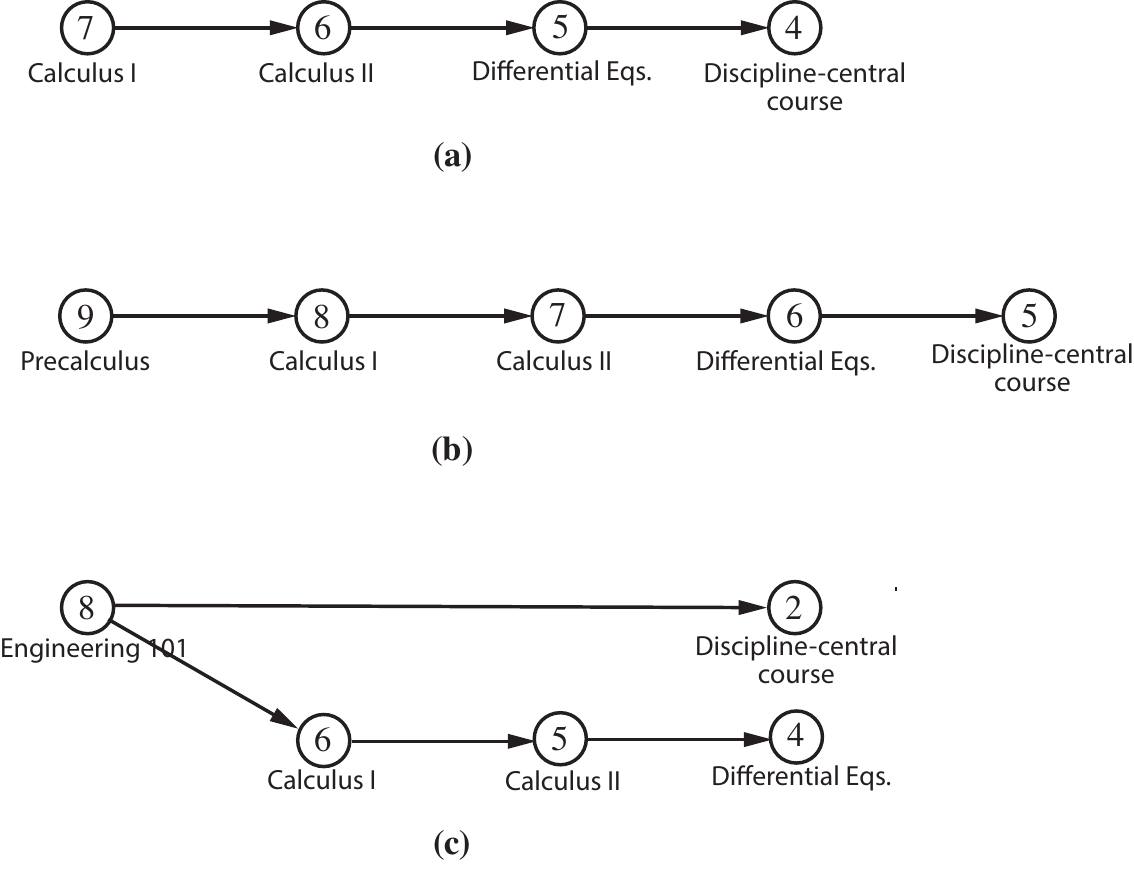}}
 \caption{Curricular design patterns for engineering programs, with course structural complexities shown inside each vertex. (a) A curricular design pattern found in most engineering programs constructed under the assumption that students are calculus ready. The structural complexity of the pattern is 22. (b) A modified pattern that is often used if student are not calculus ready, leading to a structural complexity of 35. (c) A revised curricular design pattern that significantly lowers the structural complexity for non-calculus-ready students to 25.}\label{EngPattern}
\end{figure}

Next we describe how previous work in the area of curriculum redesign can be cast into the realm of curricular design patterns, thereby allowing us to apply curricular analytics as a means to more formally analyze the benefits of curriculum redesign work. We consider the curriculum redesign efforts of \cite{KlBo:15}, who recognized the deleterious effects that applying the common remedy of simply prepending math prerequisites onto standard engineering curricula can have on non-calculus-ready students. They noticed that in most engineering programs there is a course central to the discipline that is typically taken in the sophomore year, and that Differential Equations serves as a prerequisite to this course. This can be expressed as the curricular design pattern shown in  Figure~\ref{EngPattern}~(a), which assumes students begin the pattern calculus ready. 
If, however, a student is not calculus ready, the common remedy applied in many STEM programs involves simply prepending a Precalculus course to the aforementioned pattern, as shown in Figure~\ref{EngPattern}~(b).\footnote{If a student is not precalculus ready, the common remedy is to prepend additional prerequisite math courses to the Precalculus course. At some institutions there may be as many as three math courses preceding Precalculus, and we have seen cases, e.g., Figure~\ref{univ-curric-db}~(a), where this would create a path in a curriculum that spans eleven terms, making it impossible for students who start at the first course in this math sequence to complete the degree in fewer than five-and-one-half years. In this case, the sophomore-level Circuits I class could not be taken until the seventh term, i.e., in what should be a student's senior year.} 

\cite{KlBo:15} also noticed that only a small portion of the learning outcomes in Differential Equations, namely the ability to solve linear differential equations, are actually used in most of the aforementioned discipline-central engineering courses. A typical Differential Equations course will devote a few weeks to these simple differential equations, prior to considering a wide range of significantly more complicated categories of differential equations. This observation provides us with the opportunity to rearrange courses within curricular patterns in ways that are beneficial to certain student populations. This is precisely what \cite{KlBo:15} did as a part of their curricular redesign efforts aimed at providing engineering curricula more suitable for students who are not calculus ready when they enter college. Their insights around the types of differential equations required in certain discipline-specific engineering courses led them to consider a redesign that involved teaching the ability to solve linear differential equations in a high-impact first-year engineering course that also includes precalculus topics. The resulting curriculum is shown in Figure~\ref{EngPattern}~(c). Notice that students are still required to take the Differential Equations course in this curriculum. Notice also that in this revised curriculum, students starting in Engineering 101 are able to matriculate to the sophomore-level discipline-specific engineering course at the same time as, or even sooner than, their classmates who begin college calculus ready.

The methodologies described in this paper can be used to quantify the benefits of the \cite{KlBo:15} approach by demonstrating the extent to which their reform reduces structural complexity. Specifically, their remedy leads to a ten point reduction in structural complexity relative to the standard remedy shown in Figure~\ref{EngPattern}~(b), and it is only three points more complex than the pattern used by calculus-ready students. To understand the impact this can have on student success in engineering programs, consider the case where the pass rates are 75\% for all courses in the design patterns shown in Figure~\ref{EngPattern}. In this case, after six terms, on average 82\% of students are able to complete the pattern in Figure~\ref{EngPattern}~(a), but only 53\% are able to complete the pattern in Figure~\ref{EngPattern}~(b). Students attempting to complete the pattern in Figure~\ref{EngPattern}~(c) within six terms will succeed  on average 83\% of the time. The additional flexibility of being able to shift the discipline-specific course in pattern Figure~\ref{EngPattern}~(c) to any of the last three terms in the pattern~(assuming a student passes Engineering 101 in the first term) actually makes this pattern roughly equivalent, in terms of success, to the patten provided to calculus-ready students in Figure~\ref{EngPattern}~(c). Furthermore, one of the main motivations behind creating a high-impact first-year course in the major is to increase student success, typically by ``layering on''  additional support services. If these support services are effective, it is reasonable to assume the pass rate of Engineering 101 would increase to 95\%, in which case the success rate for students attempting the pattern in Figure~\ref{EngPattern}~(c) jumps to 88\%.
 
In order to better understand how one might apply the curricular design pattern suggested in Figure~\ref{EngPattern}~(c), we will consider the electrical engineering context shown in Figure~\ref{univ-curric}. We previously noted the centrality of Circuits~I in electrical engineering curricula.  The learning outcomes students should attain in Circuits I include the ability to:
\begin{enumerate}
 \item Understand the functions of basic electrical circuit elements and sources;
 \item Apply Ohm's and Kirchhoff's circuit laws in the lumped element model of electrical circuits;
 \item Appreciate the consequences of linearity, in particular the principle of superposition and Thevenin and Norton equivalent circuits;
 \item Understand the concept of state in a dynamical physical system and analyze simple first and second order linear circuits containing memory elements.
\end{enumerate}

\begin{figure}
  \centerline{\includegraphics{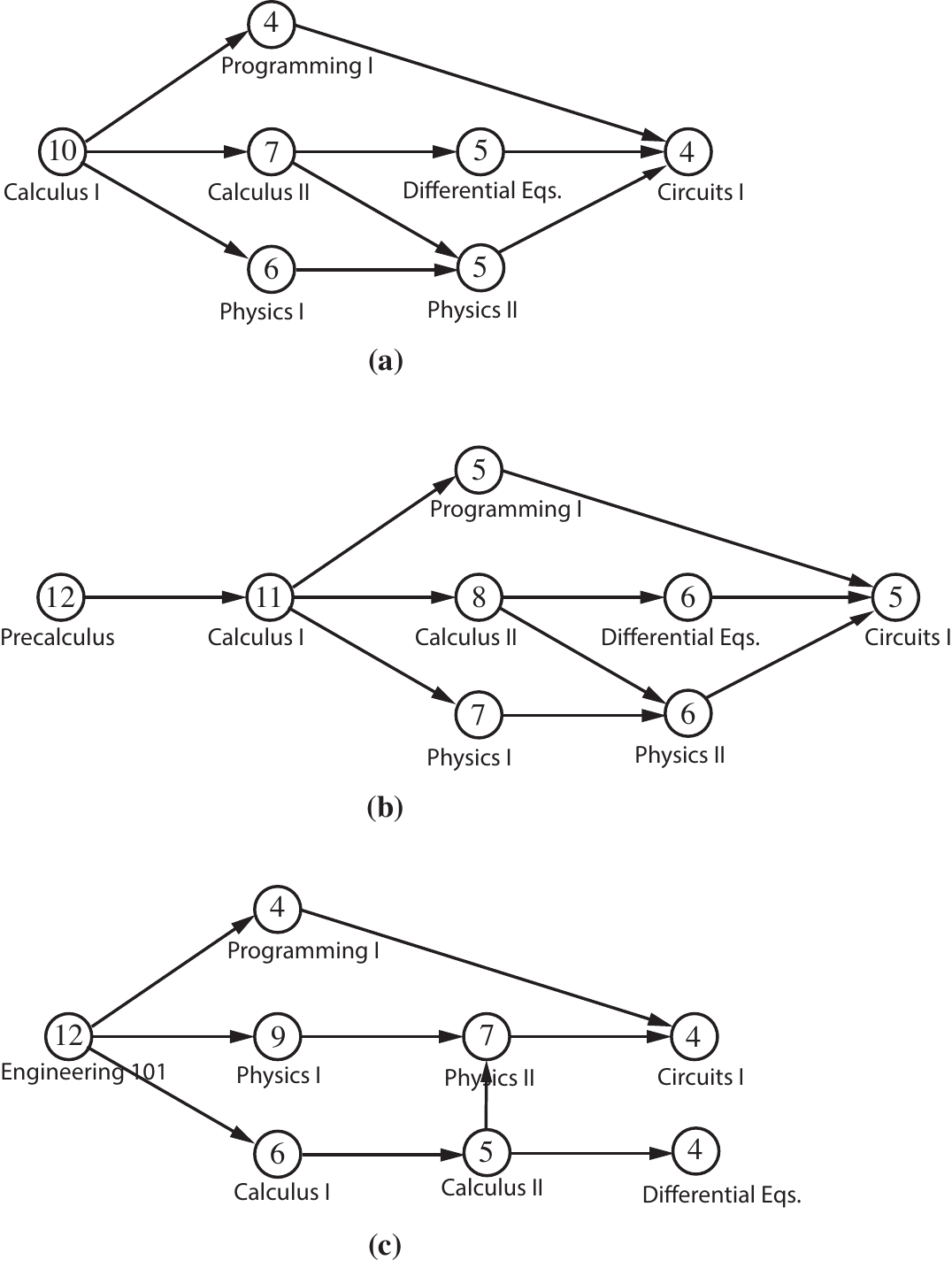}}
 \caption{Instantiations of the curricular design patterns shown in Figure~\ref{EngPattern} in the electrical engineering context that allow for the attainment of the Circuits I learning outcomes. (a) A four-term design pattern for students who are Calculus I ready, with an overall structural complexity of 41. (b) A five-term design pattern for students who are not Calculus I ready, with an overall structural complexity of 60. (c) A four-term design pattern for students who are not Calculus I ready, with an overall structural complexity of  51.}\label{CircuitsI}
\end{figure}

A seven-course instantiation of the curricular pattern shown in Figure~\ref{EngPattern}~(a) designed to allow students to attain the Circuits I learning outcomes, under the assumption that students are calculus ready, is shown in Figure~\ref{CircuitsI}~(a). 
Inside of each vertex in this figure we show the course structural complexity computed using Equation~(\ref{struct_complex_fit}). The structural complexity of the entire curricular design pattern in Figure~\ref{CircuitsI}~(a) is 41, and the longest path in this pattern includes four courses, which means the pattern cannot be completed in fewer than four terms.

In  Figure~\ref{CircuitsI}~(b) we provide an eight-course instantiation of the curricular design pattern in Figure~\ref{EngPattern}~(b) that allows students to attain the Circuits~I learning outcomes under the assumption that students are \emph{not} initially prepared for Calculus I. This pattern uses the common remedy previously alluded to, which is to tell students to first make themselves Calculus I ready, and then attempt the same curriculum offered to the calculus-ready students. That is, by applying the common remedy of prepending Precalculus, the structural complexity of the curricular design pattern in Figure~\ref{CircuitsI}~(b) is 60, a 31\% increase over the one provided in Figure~\ref{CircuitsI}~(a). Notice also that the curricular design pattern in Figure~\ref{CircuitsI}~(b) cannot be completed in fewer than five terms.

The key observation of \cite{KlBo:15} that only a subset of the learning activities in a given course may be necessary as prerequisite material for the learning activities in specific follow-on courses applies to the Circuits~I learning outcomes. Specifically, the fourth learning outcome listed above requires that a student have prior learning that includes the ability to solve first and second order linear differential equations. Thus, by applying the Figure~\ref{EngPattern}(c) curriculum design pattern in this context, we obtain the remedy shown in Figure~\ref{CircuitsI}~(c). Relative to the common remedy shown in Figure~\ref{CircuitsI}~(b), this yields a four-term curricular pattern with an overall structural complexity of~51, making the pathway for students following the Figure~\ref{CircuitsI}~(c) pattern only 20\% more complex than the standard calculus-ready pattern in Figure~\ref{CircuitsI}~(a).\footnote{In Figure~\ref{CircuitsI}~(c),  Calculus II is a co-requisite for Physics II, a common practice in higher education.} 

The curricular design pattern shown in Figure~\ref{CircuitsI}~(c) allows students to attain the same set of learning outcomes as those shown in Figures~\ref{CircuitsI}~(a) and~(b). It is similar to the design pattern in Figure~\ref{CircuitsI}~(b) in that it serves students who are not calculus ready, but it differs from that design pattern in that its structural complexity is much lower---the pattern in Figure~\ref{CircuitsI}~(c) can be completed in four terms, rather than five as in Figure~\ref{CircuitsI}~(b). Thus, we would expect students to be more successful in completing Figure~\ref{CircuitsI}~(c), as compared to Figure~\ref{CircuitsI}~(b). We can validate this through simulation. Specifically, if we fix the pass rates of all courses at 75\%, after six terms, on average 72\% of students are able to complete the pattern in Figure~\ref{CircuitsI}~(a), but only 36\% are able to complete the pattern in Figure~\ref{CircuitsI}~(b). Students attempting to complete the pattern in Figure~\ref{EngPattern}~(c) within six terms will succeed  on average 72\% of the time; that is, at the same rate as the students attempting the calculus-ready pattern. The manner in which curricular analytics can be used to derive alternative curricula, such as the one shown in Figure~\ref{CircuitsI}~(c) is considered next.

\subsection{Degree Plan Construction and Curriculum Deconstruction}\label{deconstruction}
The curricula shown in Figure~\ref{univ-curric} are organized as plans that allow students to satisfy all degree requirements in eight terms. Many other valid degree plans could be created for these degree programs. In other words, there is a one-to-many relationship between degree requirements and degree plans. A logical question to ask is, with regards to student success, are some degree plans for a given program better than others? Because the overall structural complexity of a curriculum depends only on the pre- and co-requisite relationships between courses, all valid degree plans for a given curriculum will have the same structural complexity. What varies between the degree plans for a curriculum is how this structural complexity is distributed over the terms in these plans. It is possible to develop different criteria for distributing complexity throughout a plan, and to create optimized degree plans according to these criteria. For instance, \cite{SlHeLoAlAb:15} recognized the importance of having students complete the most crucial courses in a curriculum, as measured by course complexity, as early as possible in a degree plan, and they provided an algorithm for constructing degree plans optimized according to this strategy. Another approach is to balance out the complexity over the first few years so that no one term is overly complex. This approach may be better suited to students who are more susceptible to stopping out of college when they encounter difficulties. 

The prior discussion suggests that it makes sense to select optimal degree plans on a per student basis. In other words, from a student success perspective, a useful line of research involves investigating the possibility of constructing degree plans tuned to the capabilities of individual students. One approach recognizes that some aspects of instructional complexity are conditionally dependent on the characteristics of individual students and the courses that are grouped together in a term. Thus, by creating a measure for instructional complexity that captures the expected performance of different categories of students in particular combinations of classes, we have a means for creating degree plans that optimize the likelihood of success for these categories of students. Further investigation of this approach is warranted. 

The aforementioned strategies around degree plan construction assume a fixed curriculum graph. We saw, however, in Section~\ref{design_patterns} that it is possible to create beneficial solutions by restructuring the curriculum graph itself. Indeed many curriculum reform efforts correspond to modifying the curriculum graph, even if the reformers are not aware of this technical detail. The example described in Section~\ref{design_patterns} actually suggests a more formal approach for exploring curriculum redesign through \emph{curricular decomposition}. Specifically, a redesign effort that uses curricular decomposition starts by decomposing some portion of a curriculum into the learning outcomes associated with the courses in that portion of the curriculum. The next step involves documenting the dependencies between  these learning outcomes; that is, the prerequisite structure of the learning outcomes. Finally, curriculum reformers consider the various ways that these learning outcomes can be reassembled back into courses, adding a prerequisite whenever there are learning outcome dependencies that cross course boundaries. In essence, this is a formalization of the intuitive curriculum redesign approach described in Section~\ref{design_patterns}. It is important to note that different arrangements of learning outcomes in courses will produce curricula with different structural complexities. Thus, what we have articulated is a means of automating curricular redesign that will allow us to create algorithms that can search for curricular improvements. This approach also warrants further investigation. 

%------------------------------------------------
\section{Concluding Remarks}\label{conclusion}
In this paper we presented a framework for the study of curricula that treats a curriculum as the unit analysis. By treating a curriculum as a formal system that exists within the larger university ecosystem, we have highlighted the fact that it can be directly and rigorously analyzed. An analytical approach to the study of curricula supports not only the ability to make predictions about how curricular changes will effect student progress, but also predictions around the likely impact of particular student success interventions on curricular progression. 

Our approach involved separating the factors that influence student progression through a curriculum into two independent sets of factors, one set captures the structure of a curriculum, and the other takes into account the manner in which the courses in a curriculum are supported and taught. We derived a novel set of useful analytics from this formulation that allowed us to directly quantify the impact of curricular factors on student success outcomes. Finally, we considered a number of different practical setting where this type of analysis may be usefully applied. In our view, we have only scratched the surface on the use of curricular analytics as a tool for guiding student success interventions, and in Section~\ref{application} we alluded to a few ways this work can be extended.

The main contribution of this work is the broad analytical framework it provides for the direct study of curricula. As with any useful theory created to model a physical or sociological phenomenon, application of the theory provides an approximation of the behavior one would expect to observe when the underlying system is subjected to various internal or external influences. We are well aware of the fact that differences in how various institutions report their curricula can lead to small differences in the complexity metrics we have derived. Consider, for instance, a science course with an associated laboratory section that is treated as a single four-credit course in one curriculum, but as a three-credit lecture section combined with a one-credit laboratory, with a co-requisite between them, in another curriculum. Within the context of an entire curriculum, the latter will appear as slightly more complex than the former when using our structural complexity metrics. Similarly, there may be prerequisites that are not strongly enforced by program administrators, or prerequisite arrangements that are implied without being explicitly noted in a curriculum graph. As with any theory, it is the reasoned application by skilled practitioners of the methodologies derived from the theory that should guide work and interpretation of results. The aforementioned scenarios could easily be incorporated into a revised structural complexity metric if their influence was believed to be significant. It is our experience, however, that the manner in which these scenarios are treated tend to only slightly perturb the overall complexity measure of an entire curricular system. They may, however, provide a significant influence on a small portion of a curriculum that is considered problematic. In this case, a more accurate accounting of the particular curricular details may be important. Again, it is up to the practitioner to apply the appropriate tool from the available theoretical toolset.

The phrase ``imitation is the sincerest form of flattery'' certainly applies to higher education, where colleges and universities over many centuries have borrowed successful practices from one another, often as a means of mimicking highly regarded institutions. Organizational structure, governance, pedagogy and course offerings are remarkably similar across the various sectors of higher education. We learn from our colleagues, and we try to replicate what works. 
One challenge associated with further extending the efficacy of curricular analytics involves facilitating the ability to compare similar curricula at different institutions. We provided one such comparison in Figure~\ref{univ-curric}; however, this necessitated a significant amount of investigation into the course descriptions provided on university websites in order to similarly name the courses listed in these two curricula. Those within a given discipline typically share a vocabulary that they use to describe the courses commonly taught in their curricula. We refer to these as the \emph{canonical course names} for a discipline.  In Figure~\ref{univ-curric}, to the extent possible we used canonical course names, rather than university-specific names, e.g., Calculus~I rather than Math 192. 
A curriculum database built using canonical course names would support the ability to compare and contrast curricula in an automated fashion. For instance, it would enable the ability to search through collections of curriculum graphs in order to find common or anomalous patterns within disciplines. We contend that the creation of a catalog of common patterns, as well as anti-patterns, would greatly facilitate curriculum reform efforts, and also ground discussions within a formal framework.\footnote{Anti-patterns are common or obvious solutions to recurring design problems that have proven ineffective in practice.} Furthermore, by incorporating historical student success data, we would have a means for determining if certain curricular design patterns are more suitable for particular populations of students, supporting the ability to tune curricula to the needs of our students.  

%----------------------------------------------------------------------------------------
\bibliography{academic} 
\bibliographystyle{apalike} 

\end{document}